\documentclass[preprint2,tigten,letteredappendix,appendixfloats]{aastex63}
\input{emulateapj_look_and_feel.tex}
\usepackage[strict]{changepage}
\newenvironment{ctabular}[1]{\begin{adjustwidth}{-2.82in}{-2in}\centering\begin{tabular}{#1}}{\end{tabular}\end{adjustwidth}}

\usepackage{afterpage}
\usepackage{natbib}
\usepackage{enumitem}
\usepackage{amsmath}
\usepackage{soul}
\usepackage[all]{hypcap}

\let\tablenum\relax\usepackage[load-configurations=astronomy]{siunitx}
\DeclareSIUnit\micron{\ensuremath{\mathrm{\micro\meter}}}
\DeclareSIUnit\sig{\ensuremath{\mathrm{\sigma}}}
\DeclareSIUnit\mag{mag}
\DeclareSIUnit\jansky{Jy}
\DeclareSIUnit\deg{deg}
\DeclareSIUnit\yr{yr}
\DeclareSIUnit\dex{dex}
\DeclareSIUnit\Lsun{\ensuremath{\mathrm{L_\odot}}}
\DeclareSIUnit\Msun{\ensuremath{\mathrm{M_\sun}}}

\newcommand{\midrule}{\vspace{-2.0ex}\\\hline\vspace{-1.5ex}\\}
\newcommand{\bottomrule}{\vspace{-2.0ex}\\\hline\vspace{-1.5ex}\\}

\newcommand{\Herschel}{\textit{Herschel}}
\newcommand{\HIPE}{\texttt{HIPE}}

\begin{document}

\title{A Systematic Search for the Reddest Far-infrared and Sub-millimeter 
Galaxies: revealing dust-embedded starbursts at high redshifts}

\author{Haojing Yan}\affiliation{Department of Physics and Astronomy, University of Missouri-Columbia, USA}
\email{yanha@missouri.edu}

\author{Zhiyuan Ma}
\affiliation{Department of Astronomy, University of Massachusetts, Amherst, USA}
\email{zhiyuanma@umass.edu}

\author{Jia-Sheng Huang}\affiliation{CASSACA, National Astronomical Observatories, Chinese Academy
of Sciences, Beijing 100012, China}
\email{jhuang@nao.cas.cn}

\author{Lulu Fan}\affiliation{CAS Key Laboratory for Research in Galaxies and Cosmology, Department of Astronomy, University of Science and Technology of China, Hefei 230026, China}
\affiliation{School of Astronomy and Space Sciences, University of Science and Technology of China, Hefei, Anhui 230026, China}
\affiliation{Shandong Provincial Key Lab of Optical Astronomy and Solar-Terrestrial Environment, Institute of Space Science, Shandong University, Weihai, 264209, China}
\email{llfan@ustc.edu.cn}

\begin{abstract}

  We present the results of our systematic search for the reddest far-infrared
(FIR) and submillimeter (sub-mm) galaxies using the data from the Herschel
Multi-tiered Extragalactic Survey (HerMES) and the SCUBA2 Cosmological Legacy
Survey (S2CLS). The red FIR galaxies are ``500~$\mu$m risers,'' whose spectral
energy distributions increase with wavelength across the three FIR
passbands of the Spectral and Photometric Imaging REceiver (SPIRE) of
Herschel. Within 106.5~deg$^2$ of the HerMES fields, we have selected 629
500~$\mu$m risers. The red sub-mm galaxies are
``SPIRE dropouts,'' which are prominent detections in the S2CLS 850~$\mu$m data
but are extremely weak or invisible in the SPIRE bands. Within the 2.98~deg$^2$
common area of HerMES and S2CLS, we have selected 95 such objects. These very
red sources could be dusty starbursts at high redshifts ($z\gtrsim 4$--6)
because the peak of their cold-dust emission heated by star formation is
shifted to the reddest FIR/sub-mm bands. The surface density of 500~$\mu$m
risers is $\sim$8.2~deg$^{-2}$ at the $\geq 20$~mJy level in 500~$\mu$m, while
that of SPIRE dropouts is $\sim$19.3~deg$^{-2}$ at the $\geq 5$~mJy level in
850~$\mu$m. Both type of objects could span a wide range of redshifts, however. Using deep radio data in these fields to further select the ones likely at the
highest redshifts, we find that the surface density of $z>6$ candidates is 
5.5~deg$^{-2}$ among 500~$\mu$m risers and is 0.8--13.6~deg$^{-2}$ among SPIRE
dropouts. If this is correct, the dust-embedded star formation processes in 
such objects could contribute comparably as Lyman-break galaxies to the global
SFR density at $z>6$.

\end{abstract}


\section{Introduction}

   A surprising fact that has recently emerged from the exploration of the 
early universe is that very dusty starburst galaxies exist at $z>6.3$, which is
in the epoch of cosmic hydrogen reionization
\citep[EoR; e.g.,][]{Fan2006, Planck2016reion}.
Some of these objects were discovered using quasars at $z\gtrsim 6$--7 as 
probes, revealed as either being their host galaxies
\citep[e.g.,][]{Walter2003, Venemans2012, Venemans2016, Venemans2017},
or being their companions \citep[][]{Decarli2017},
and some were discovered by targeting sources found in far-infrared (FIR) to
submillimeter/millimeter (sub-mm/mm) surveys
\citep[][]{Riechers2013, Strandet2017}.
Their IR luminosities ($L_{IR}$; integrated over rest-frame 8--1000~$\mu$m)
are so high that they qualify as ultraluminous infrared galaxies (ULIRGs; 
$L_{IR}\geq 10^{12}L_\odot$), and the inferred star formation rates (SFRs) are
often an order of magnitude higher than those of the most luminous galaxies
at the same redshifts found through the conventional ``dropout'' technique
based on rest-frame UV emission.  Therefore, there is a possibility that 
dust-embedded star formation activities in the universe could exceed those 
``exposed'' ones even in such an early epoch. Most of these currently known
dusty starbursts in the EoR are invisible in the deep rest-frame UV images,
and thus their dust-hidden star formation cannot be recovered through the
extinction-corrected UV emission. Their huge amount of dust suggests
very active star formation in even earlier epochs and a short
time scale of metal enrichment throughout the ISM, both of which
remain to be verified by future observations.
This population also impacts our understanding of the intergalactic medium
(IGM) in the early universe.
The necessary condition for the reionization to happen is a
strong Lyman photon background, which is already difficult to account for
unless the UV luminosity functions of galaxies at $z>6$ have very steep
faint-end slopes \citep[e.g.,][]{Yan2004a, Yan2010, Bouwens2015}.
The existence of dusty starbursts at $z>6$, albeit only a small number having
been confirmed to date, indicates that dust could already be ubiquitous in 
galaxies when the universe was merely a few hundred million years old.
This would exacerbate the tight situation in explaining the Lyman photon budget
for reionization, as dust blocks UV light and thus would significantly decrease
the Lyman photon escape fraction. 

   Currently, the sample of confirmed dusty starbursts in the EoR is still 
very small. The most promising, unbiased way (i.e., without being tied to
the known high-$z$ quasars) to significantly increase the sample size is to do 
sub-mm/mm spectroscopy on large sets of candidates preselected based on their
spectral energy distributions (SEDs) in the FIR-to-mm regime. Such a 
preselection has recently been made possible by the wide-field FIR surveys
done using the Spectral and Photometric Imaging Receiver
\citep[SPIRE;][]{Griffin2010} onboard the Herschel Space Observatory
\citep[][]{Pilbratt2010}, which offered high-sensitivity photometry in its three
broad bands at 250, 350, and 500~$\mu$m. \Herschel\, completed two major
extragalactic surveys with this instrument, namely the Herschel Astrophysical
Terahertz Large Area Survey \citep[H-ATLAS;][]{Eales2010, Smith2017} over
$\sim$660~deg$^2$ in five areas and the Herschel Multi-tiered Extragalactic
Survey \citep[HerMES;][]{Oliver2012} over $\sim$70~deg$^2$ in six tiered levels
of depth and spatial coverage combinations (the released data actually cover
wider areas). In addition, HerMES extended beyond
its shallowest tier to the Herschel Stripe 82 Survey \citep[HerS;][]{Viero2014}
and the HerMES Large Mode Survey \citep[HeLMS; see][]{Asboth2016}, which
observed $\sim 80$ and $\sim 280$~deg$^2$ in and around the Sloan Digital Sky
Survey Stripe 82 region, respectively. Combining all of these surveys, there 
are now over a million FIR sources awaiting exploration.

   Tailored for the SPIRE bands, the so-called ``500~$\mu$m peaker'' or 
``500~$\mu$m riser'' technique selects candidates of high-redshift ULIRGs
by searching for sources that are progressively brighter from 250 to 500~$\mu$m
\citep[][]{Pope2010, Roseboom2012}. This is based on the fact that young stars
nominally can only heat dust to a few tens of kelvin, and therefore the 
FIR/sub-mm SED can be approximated well by a modified blackbody (MBB) emission
that has a broad peak at roughly 80--100~$\mu$m. Within the SPIRE bands, this 
peak would move to the redder side of 250~$\mu$m at longer wavelengths if the
object is at $z\gtrsim 4$, and hence the SED shows a rising trend from 250 to 
500~$\mu$m. This method has directly resulted in the ULIRG at $z=6.337$ 
presented by \citet[][]{Riechers2013}, which was then the redshift record
holder among all dusty starbursts that are not related to quasars. The current
record holder in this category, which is at $z=6.900$ 
\citep[][selected as a red mm source]{Strandet2017}, 
is also a 500~$\mu$m riser based on its SPIRE photometry obtained after the 
source redshift was identified. Admittedly, this method has the drawback that
it does not have sufficient resolution power in redshift; instead, it can only
select sources at $z\gtrsim 4$ in general. This is due to the degeneracy
between dust temperature ($T_d$) and redshift \citep[see, e.g.][]{Pope2010}: a
lower $T_d$ and a fixed redshift would shift the peak of the SED to a longer
wavelength, which would be the same behavior if $T_d$ is kept the same but the
source is moved to a higher redshift. Regardless, this method is still the most
effective in producing a large number of legitimate candidates at $z\gtrsim 4$
in a comprehensive manner, and the samples at different redshift ranges will
need to be separated after spectroscopic identification. The latter is 
particularly important for the study of objects in the EoR, as they might 
constitute only a small fraction among the candidates thus selected.

   Using the early HerMES data, \citet[][]{Dowell2014} presented the first 
systematic search for 500~$\mu$m risers in three fields (First Look Survey,
Lockman Hole, and GOODS-North) totaling 21~deg$^2$, which resulted in 38
objects. They used the ``difference map'' technique, where the candidate
selection is done on the difference map between the 500 and the 250~$\mu$m maps
after smoothing the latter to the beam size of the former (i.e., from 
17.\arcsec6 to 35.\arcsec2 FWHM) and assigning some certain weights to each.
\citet[][]{Asboth2016} applied the same method to the 274~deg$^2$ HeLMS field
and published 477 500~$\mu$m risers. \citet[][]{Ivison2016} took a different
approach in the 660~deg$^2$ H-ATLAS fields. These authors started from the goal
of constructing the most complete source catalog: sources are first identified
on the 250~$\mu$m map (which has the best solution and sensitivity) and removed
from the 350 and the 500$\mu$m maps, additional sources are identified on these
350 and 500$~\mu$m residual maps, and candidate 500~$\mu$m risers are then
selected from this complete catalog. They obtained 7961 500~$\mu$m risers in
total, at least $\sim$$26\pm 5$\% of which should be reliable. 
\citet[][]{Donevski2018} used the similar methodology (but a different
implementation) to search for 500~$\mu$m risers in the 55~deg$^2$ Herschel Virgo
Cluster Survey, and selected 133 such objects.

   An extension of the 500~$\mu$m riser technique to longer wavelengths has 
recently been proposed, which selects objects that continue the rising
SED trend to the red side of 500~$\mu$m. Depending on the exact location of the
redder band in use, it is referred to as the technique of ``850~$\mu$m'' or
``870~$\mu$m'' riser, etc. The dust emission peak of 850/870~$\mu$m risers 
should be at $\lambda_{obs}\gtrsim 850$/870~$\mu$m, and thus statistically, they
should be at higher redshifts than 500~$\mu$m risers. The first such object 
with confirmed redshift was presented by \citet[][]{Riechers2017}, which is an
870~$\mu$m riser and is at $z=5.655$. \citet[][]{Ivison2016} presented 850 
and/or 870~$\mu$m observations for 109 of their 7961 500~$\mu$m risers, and
only about three of them qualify as significant 850/870~$\mu$m risers
(measured using 60\arcsec\ aperture and with $\mathrm{S/N\geq5}$).
\citet[][]{Duivenvoorden2018} carried out 850~$\mu$m follow-up observations of
188 500~$\mu$m risers of \citet[][]{Asboth2016} (67 are detected at
$\mathrm{S/N\geq 5}$), however none of them qualify as 850~$\mu$m risers
(regardless of $\mathrm{S/N}$). These results suggest that such objects could
be very rare.

   In this work, we present our search of 500 and 850~$\mu$m risers using the
data from the HerMES program and the SCUBA2 Cosmological Legacy Survey 
\citep[S2CLS;][]{Geach2017, Michalowski2017}. As compared to the existing
search of 500~$\mu$m risers, ours is done at a fainter level than those by 
\citet[][]{Asboth2016} and \citet[]{Ivison2016}, and is over much large areas 
than those by \citet[][]{Dowell2014} and \citet[][]{Donevski2018}. As explained
in detail later, we do not directly select 850~$\mu$m risers due to the 
limitation of the data. Instead, we select ``SPIRE dropouts,'' which are 
objects prominent in 850~$\mu$m but very weak or invisible in SPIRE, as 
potential 850~$\mu$m risers. A similar term, ``Herschel drop-out'', was first
used by \citet[][]{Boone2013} when reporting a gravitationally lensed 
870~$\mu$m source that is invisible in Herschel data (including those obtained
by SPIRE). Their follow-up study suggests a significant population of such 
objects in lensed fields \citep[][]{Boone2015}, however the results are yet to
be published. The next explicit reference to SPIRE dropout is in the recent
paper by \citet[][]{Greenslade2019}, who present a serendipitous discovery of
such an object in the H-ATLAS North Galactic Pole field. Our selection, on the
other hand, is done in a systematic manner over the largest area available to
date.

   This paper is organized as follows. We first briefly describe the HerMES and
the S2CLS data in \S 2. The selections of 500~$\mu$m risers and SPIRE dropouts
are detailed in \S 3 and \S 4, respectively. A source-by-source comparison to
the known objects in the literature is given in \S 5. The radio properties
of both types of objects are presented in \S 6. In \S 7, we discuss their
surface densities, their prospects of being at high-$z$, and the implication of
dust-embedded star formation in the early universe. Finally, we summarize our
results in \S 8. The full catalogs of both types of objects as well as the 
ancillary catalogs are released with this paper. In our discussion, we adopt
the following cosmological parameters:
$\Omega_M=0.27$, $\Omega_\Lambda=0.73$, and $H_0=71$~km~s$^{-1}$~Mpc$^{-1}$.

\section{Primary Data Sets}

  Here we describe the data used in our selection of high-$z$ ULIRG candidates,
which are those from the HerMES and the S2CLS programs. Some other
ancillary data that we use in our analysis of the candidates will be
described in the relevant sections separately because such data are
different in different fields.

\subsection{HerMES FIR data}

   We used the SPIRE maps and the source catalogs included in the fourth data 
release (``DR4'') of the HerMES program \footnote{Released in 2016 July; see
\url{https://hedam.lam.fr/HerMES/index/dr4}}.
These maps were created using the algorithm as described in
\citet[][]{Levenson2010} and \citet[][]{Viero2013}, and incorporated all the
HerMES SPIRE observations as well as the latest calibration updates. The 
three-band source catalogs were extracted based on the ``blind'' 250~$\mu$m 
positions (referred to as ``xID250''), which were done in the same way as in 
the earlier releases \citep[][]{Wang2014}. We did not use the PACS data, as
these are not sensitive enough for our purpose.

   The fields used in our study are listed in Table 1. These do not include any
of the HerMES cluster fields, nor the two new fields that are not in the 
nominal HerMES program, namely SA13 (0.17 deg$^2$) and XMM-13hr (0.56 deg$^2$).
Furthermore, we had to exclude the Bo\"otes field because its DR4 xID 
catalog does not have 500~$\mu$m photometry due to a software problem whose 
nature is still not yet clear
\footnote{This has been a persistent problem in this field since the previous
HerMES data release (DR3), which is noted in the DR3 documentation. Basically,
the photometric code running on its 500~$\mu$m map returns all zero values for
the flux densities.}.
The field sizes listed in the table were calculated by counting the number of
pixels that have values $>1.0$ in the coverage maps as included in the third 
extension (``EXP'') of the released maps. In total, these data cover
106.54~deg$^2$. 

\subsection{S2CLS}

   As one of the JCMT Legacy Surveys, S2CLS was designed to do deep SCUBA2
mapping of eight well-studied extragalactic fields, mostly in 850~$\mu$m 
\citep[][]{Geach2017}. The 850~$\mu$m maps and the source catalogs have been
released to the public, and we refer the reader to \citet[][]{Geach2017} for
details. Five of these fields are within the HerMES coverage and hence are 
relevant to this work, which we list in Table \ref{tab:SD850summary}.
The listed field sizes were calculated by counting the pixels that are not 
assigned ``NAN'' in the RMS maps
\footnote{The released S2CLS maps in the COSMOS field actually 
incorporate some observations from a sequel program ``S2-COSMOS''
\citep[][]{Simpson2019}; however, the S2CLS source catalog in this field is
still largely confined to the footprint covered by the S2CLS when it ceased
operation ($\sim 1.35$~deg$^2$).
}.
In total, these five fields cover 2.98~deg$^2$.

\section{Selection of 500~$\mu$m risers}

   By definition, 500~$\mu$m risers are the strongest in 500~$\mu$m and become
successively weaker in 350 and 250~$\mu$m, and thus ideally, the selection
should be done using a source catalog whose detection is based on the
500~$\mu$m map. However, as the 500~$\mu$m band has the worst spatial resolution
(FWHM beam size of 36\arcsec) and hence the worst blending problem, a
500~$\mu$m-based source extraction would often result in meaningless color
measurements  because a seemingly single 500~$\mu$m source
might well be a blend of two or more sources at shorter wavelengths.
Therefore, we took a two-step approach. First, we used the HerMES xID catalogs,
which are based on the 250~$\mu$m detections and take advantage of the best
spatial resolution available (FWHM beam size of 18\arcsec) to search for
500~$\mu$m risers that are still visible in 250~$\mu$m. We then subtracted
off all the sources at the reported xID 250~$\mu$m positions from the 500~$\mu$m
maps and extracted sources again on the residual 500~$\mu$m maps to find
those 500~$\mu$m sources that are too faint to enter the 250~$\mu$m-based
catalogs. Our procedures are detailed below.

\subsection{Selection Based on HerMES DR4 xID Catalogs}

   The first pass was to use the xID catalogs to select 500~$\mu$m risers that
satisfy these criteria:
\begin{equation}\label{eq:500riser}
        f_{500}/et_{500}\geq 5\,\, and\,\, f_{500}>f_{350}>f_{250}.
\end{equation}
Here we adopt the naming convention of the HerMES DR4, where $f_{250}$,
$f_{350}$ and $f_{500}$ are the flux densities in 250, 350 and 500~$\mu$m,
respectively, while $et_{250}$, $et_{350}$ and $et_{500}$ are the total errors
in these three bands, respectively (including
the contributions from both the instrumental noise and the confusion noise).
In other words, the above criteria select objects that have S/N $\geq 5$ in
500~$\mu$m, and have successively increasing SEDs from blue to red.
The latter is effectively the same as that of \citet[][]{Dowell2014} and
\citet[][]{Asboth2016}, because these authors search for positive signals on
their difference maps (see \S 1). This is also similar to that used by
\citet[][]{Donevski2018}, with the difference that they also require
$f_{250}>13.2$~mJy and $f_{500}>30$~mJy.
We note that \citet[][]{Ivison2016} use different color criteria that require
$f_{500}/f_{350}\geq 0.85$ and $f_{500}/f_{250}\geq 1.5$.

\subsection{Selection Based on 500~$\mu$m residual maps}

   In the second pass, we used the Herschel Interactive Processing 
Environment \citep[HIPE{};][]{Ott2010}, which is the standard software package
of Herschel for handling its data, to create the base images for the analysis.
For a given field, we ran the \HIPE\, routine \texttt{subtractedFromImage} to 
subtract off all the sources in the xID catalog, regardless of their reported
S/N (some have S/N as low as $\sim 1$ in 250~$\mu$m). This left residual
images in all three SPIRE bands.

   We then used the 500~$\mu$m residual image to detect sources and to obtain
photometry in all three bands simultaneously, following the general approach as
described in \citet[][]{Wang2014}. We first ran the \texttt{StarFinder} 
software \citep[][]{Diolaiti2000} to extract sources, with the two
critical parameters \texttt{correlation\_threshold}, set to 0.9, and 
\texttt{min\_sourcedist}, set to 1. The list of sources were then
passed to the \HIPE{} task \texttt{sourceExtractorSimultaneous} to obtain the 
fluxes and the detection errors of the sources in all three bands on the
residual maps. \HIPE\, does not produce total errors that include the
confusion noise, which we had to derive based on the HerMES xID catalog of the
field. The catalog reports both the detection error (due to the instrumental
noise alone) and the total error for a given source, the latter of which is
computed by adding the terms due to the detection error and the confusion noise
error in quadrature. This allowed us to calculate the confusion noise error 
term for all the sources. We plotted the histogram of this term in each band
and adopted the peak value as the overall confusion noise in this band for all
the new sources that we found. 

   A complication that we had to deal with was that the residual 250~$\mu$m
map is not completely clean because the subtraction can never be perfect. The
``leftover'' of a bright 250~$\mu$m source on the residual 250~$\mu$m map, 
while not included in the new 500~$\mu$m source list as described above, can
still skew the flux measurement of a newly found 500~$\mu$m source in the 
simultaneous three-band photometry step if it is too close to the position of
the new 500~$\mu$m source in question. We did not have a better choice other
than discarding the affected 500~$\mu$m sources completely, which was done in a
``cleaning'' step. We ran run the \HIPE\, task 
\texttt{sourceExtractorSussextractor} with \texttt{detThreshold}=3 on the 
250~$\mu$m residual map to identify any significant leftover signatures, and 
then removed all the sources in our 500~$\mu$m residual-map-based catalog that
are within 36.\arcsec15 of any of the 250~$\mu$m leftover signatures
\footnote{In theory, the ``leftover'' of a bright
350~$\mu$m source would create the same problem, but in practice we found that
this did not need to be treated separately. This is because the 350~$\mu$m band
is much less sensitive than the 250~$\mu$m band, and the vast majority of
the 350~$\mu$m leftover signatures are already included in the 250~$\mu$m
cases.}.

   The additional 500~$\mu$m risers in a given field were then selected from
the ``cleaned'' catalog using the same criteria as in Equation
\ref{eq:500riser}. 

\begin{table*}
\caption{Summary of 500~$\mu$m riser Tier 1 sample
    \label{tab:500Rsummary}}
\hspace*{-6em}\begin{ctabular}{lrcrcrcrcc}
\toprule
HerMES Field & Area & $\mathrm{N^{500\mu m}_{conf}}$ & $\mathrm{N_x}$ & 
$\mathrm{\widetilde{f}_{500}^x}$ & $\mathrm{N_R}$ & 
$\mathrm{\widetilde{f}_{500}^R}$ & $\mathrm{N_{tot}}$ & 
$\mathrm{\widetilde{f}_{500}}$ & $\Sigma$ \\
         & 
\multicolumn{1}{c}{(\si{\deg\squared})} &
\multicolumn{1}{c}{(\si{\milli\jansky})} &
         &
\multicolumn{1}{c}{(\si{\milli\jansky})} &
         &
\multicolumn{1}{c}{(\si{\milli\jansky})} &
         &
\multicolumn{1}{c}{(\si{\milli\jansky})} &
\multicolumn{1}{c}{(\si{\per\deg\squared})} \\
\midrule
ADFS     &  7.93 & 4.72 &  24 & 37.1 & 16 & 34.5 &  40 & 35.2 &  5.04 \\ 
CDFS     & 12.57 & 3.63 & 123 & 27.9 & 17 & 20.9 & 140 & 27.6 & 11.14 \\ 
COSMOS   &  4.76 & 4.90 &  25 & 30.8 &  0 &  ... &  25 & 30.8 &  5.25 \\ 
EGS      &  3.30 & 4.05 &  24 & 27.4 &  2 & 32.9 &  26 & 28.3 &  7.88 \\ 
ELAIS-N1 & 12.71 & 4.70 &  45 & 35.0 &  6 & 43.0 &  51 & 35.7 &  4.01 \\ 
ELAIS-N2 &  8.29 & 5.10 &  25 & 38.2 & 14 & 37.3 &  39 & 37.3 &  4.70 \\ 
ELAIS-S1 &  8.46 & 4.30 &  50 & 29.5 & 18 & 29.9 &  68 & 29.7 &  8.04 \\ 
FLS      &  6.91 & 5.10 &  31 & 38.8 & 10 & 39.3 &  41 & 38.8 &  5.93 \\ 
GOODS-N  &  0.63 & 3.31 &   6 & 25.8 &  0 &  ... &   6 & 25.8 &  9.52 \\ 
Lockman  & 20.31 & 4.30 & 103 & 30.5 & 38 & 33.4 & 141 & 30.7 &  6.94 \\ 
XMM-LSS  & 20.67 & 6.05 &  31 & 42.2 & 21 & 39.7 &  52 & 41.3 &  2.52 \\ 
\bottomrule
\end{ctabular}
\tablecomments{$\mathrm{N^{500\mu m}_{conf}}$ is the confusion noise of the
500~$\mu$m map, which we use as a proxy to the sensitivity of the map.
$\mathrm{N_x}$ and
$\mathrm{N_R}$ are the number of 500~$\mu$m risers selected based on the HerMES
DR4 xID catalog and the 500~$\mu$m residual map, respectively, while 
$\mathrm{\widetilde{f}_{500}^x}$ and $\mathrm{\widetilde{f}_{500}^R}$ are their
corresponding median $f_{500}$ value. $\mathrm{N_{tot}}$ is the total number
of objects selected in the field, and $\mathrm{\widetilde{f}_{500}}$ is their
median $f_{500}$ value. $\Sigma$ is the surface density based on the total,
which is affected by the survey limit of the field
(see \S \ref{sec:500R_density}).
}
\end{table*}

\begin{table}
\vspace{0.5cm}
\caption{Summary of SPIRE dropout Tier 1 sample
    \label{tab:SD850summary}}
\begin{ctabular}{lcccc}
\toprule
S2CLS Field  &  Area &  N  & $S_{850}^{med}$ & $\Sigma$ \\
      &    \multicolumn{1}{c}{(\si{\deg\squared})} &
      &    \multicolumn{1}{c}{(\si{\milli\jansky})} 
      &    \multicolumn{1}{c}{(\si{\per\deg\squared})} \\
\midrule
COSMOS  &  1.34 &  26 &  6.3 & 19.4 \\
EGS     &  0.32 &   6 &  5.3 & 18.8 \\
LH      &  0.28 &   3 &  6.3 & 10.7 \\
GOODS-N &  0.07 &   3 &  6.7 & 42.9 \\
UDS     &  0.94 &  57 &  4.5 & 60.6 \\
\bottomrule
\end{ctabular}
\tablecomments{N is the total number of SPIRE dropouts found in the field 
(those found by Methods A and B combined) and $S_{850}^{med}$ is the
median deboosted $S_{850}$. $\Sigma$ is the surface density, which is affected
by both the survey limit and the size of the field (see
\S \ref{sec:SD850_density}). No correction of incompleteness or contamination
is taken into account.
}
\end{table}

\subsection{Final catalogs}

\begin{figure*}[]
\plotone{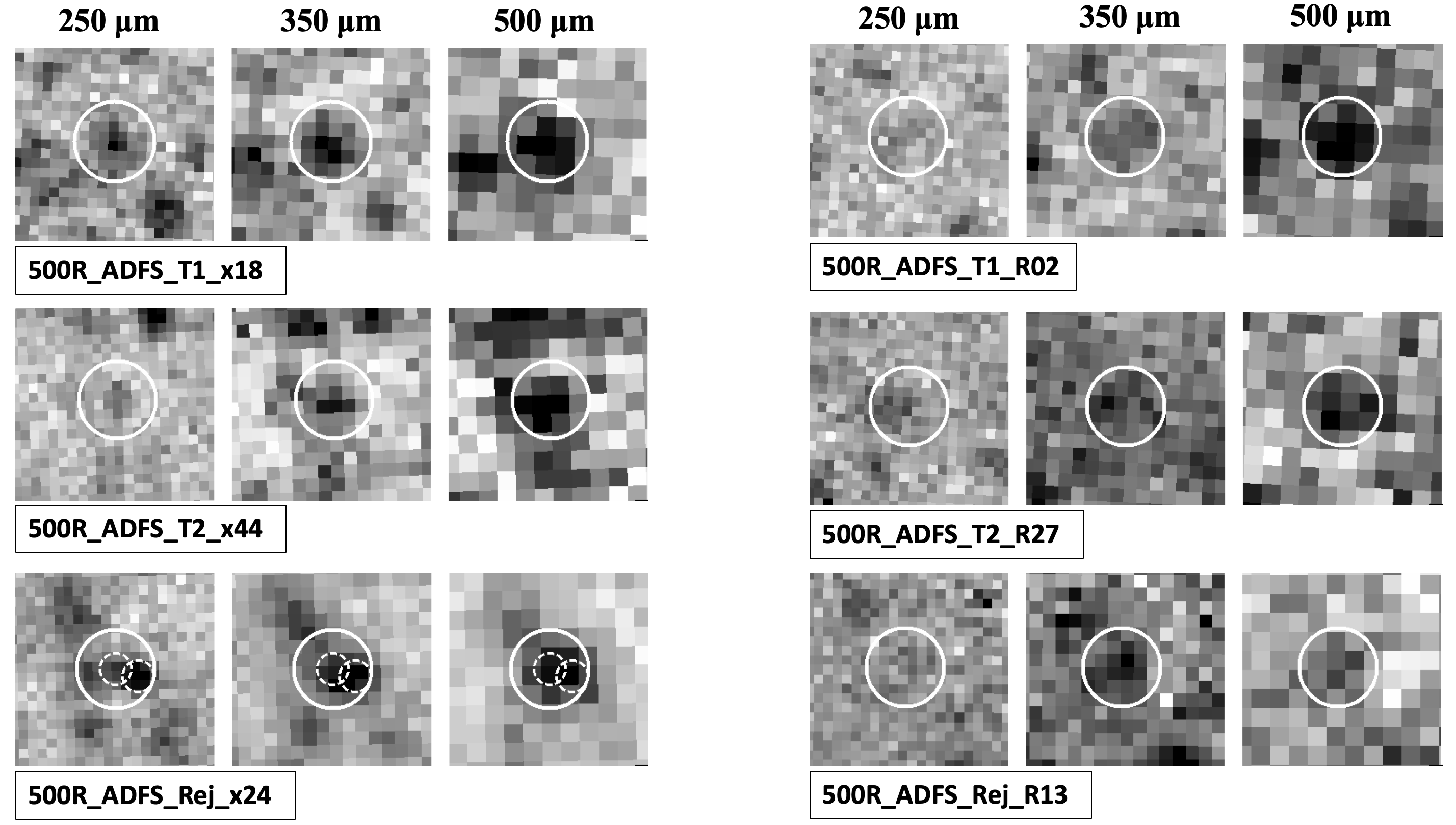}
\caption{SPIRE image stamps of some 500~$\mu$m riser candidates in the ADFS
field as examples. The ones shown on the left panel are selected using the
HerMES DR4 xID catalog, while those on the right panel are selected using the
detections based on the 500~$\mu$m residual map produced after subtracting the
250~$\mu$m-detected sources. The labels in the boxes are the object IDs. The
stamps are 2\arcmin$\times$2\arcmin\, in size and are oriented north up and
east left. The white circles are 25\arcsec\, in radius and are centered on the
candidates. In each panel, (1) the images from left to right are in 250, 350,
and 500~$\mu$m, respectively; (2) the first row is a Tier 1 object; (3) the
second row is a Tier 2 object; and (4) the last row is a rejected contaminator.
The rejected one in the left panel is due to unreliable photometry caused by
two blended objects (indicated by the two small, dashed circles), while the one
in the right panel is due to the false positive in 500~$\mu$m. 
}
\label{fig:500umRisersDemo}
\end{figure*}

   We visually examined the candidates selected above in all three SPIRE bands
to reject any possible contaminators. In most cases, the contamination is
either due to a false positive in 500~$\mu$m or due to severe blending that 
makes the photometry (especially that in 500~$\mu$m) unreliable. We have 
created our final catalogs in two tiers. In the Tier 1 catalogs, we only retain
those objects that are visually prominent in 500~$\mu$m and are the least 
affected by blending. Some of them still have close neighbors; however, we have
judged based on our experience that their photometry should still be reliable.
There are also a significant number of borderline cases that cannot be included
in the Tier 1 catalogs. For the sake of completeness, we put them to the Tier 2
catalogs. In Figure~\ref{fig:500umRisersDemo}, we show some examples of the 
objects in both tiers as well as the rejected contaminators. 
Table~\ref{tab:500Rsummary} summarizes the distribution of 500~$\mu$m risers
in all 11 fields, and the full catalogs of Tier 1 objects are presented in 
Table~\ref{tab:500RCat}. The Tier 2 objects are given in Appendix.

\begin{figure*}[]
\plotone{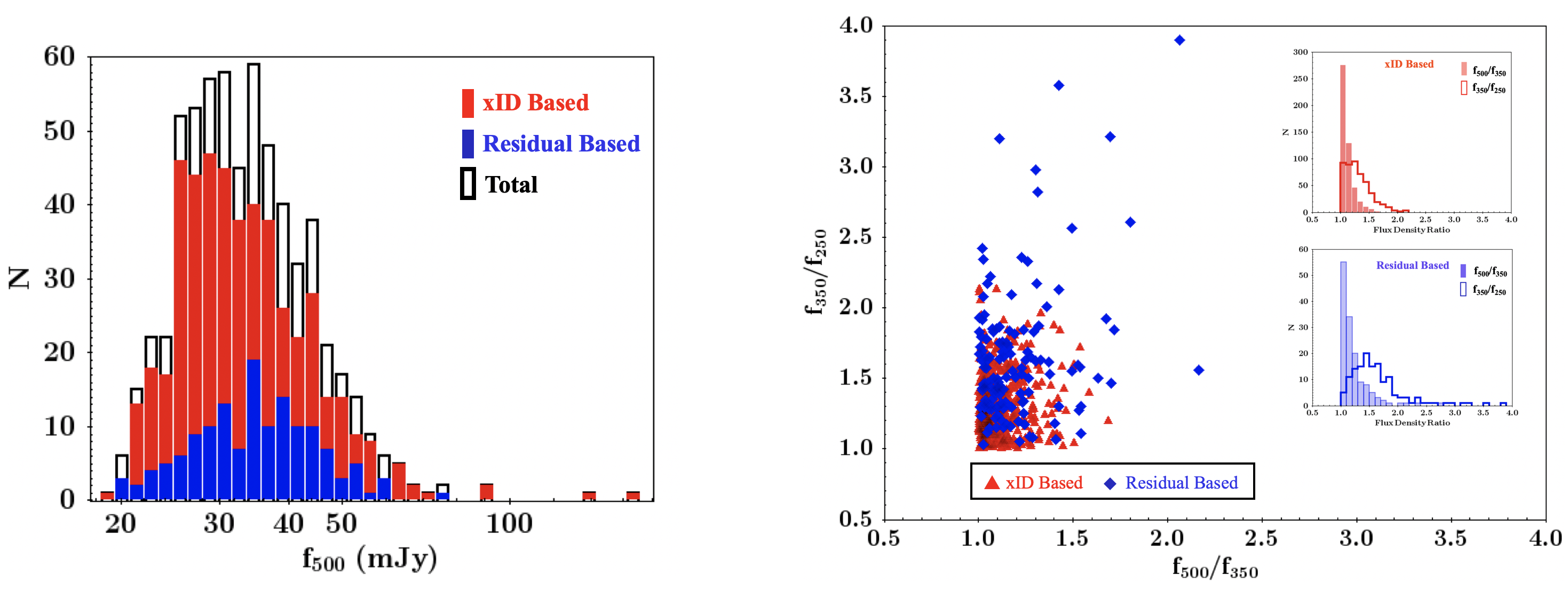}
\caption{(Left) Distributions of $f_{500}$ for the 500~$\mu$m risers selected
based on the HerMES DR4 xID catalogs (filled red) and the residual
maps (filled blue), and the total (open black).
(Right) Flux density ratios of the 500~$\mu$m risers selected based on the
HerMES DR4 xID catalogs (red triangles) and the residual maps (blue diamonds).
The insets show the histograms of the ratios for the two sets, respectively.
}
\label{fig:500umRiserFluxColor}
\end{figure*}

\begin{figure*}[]
\plotone{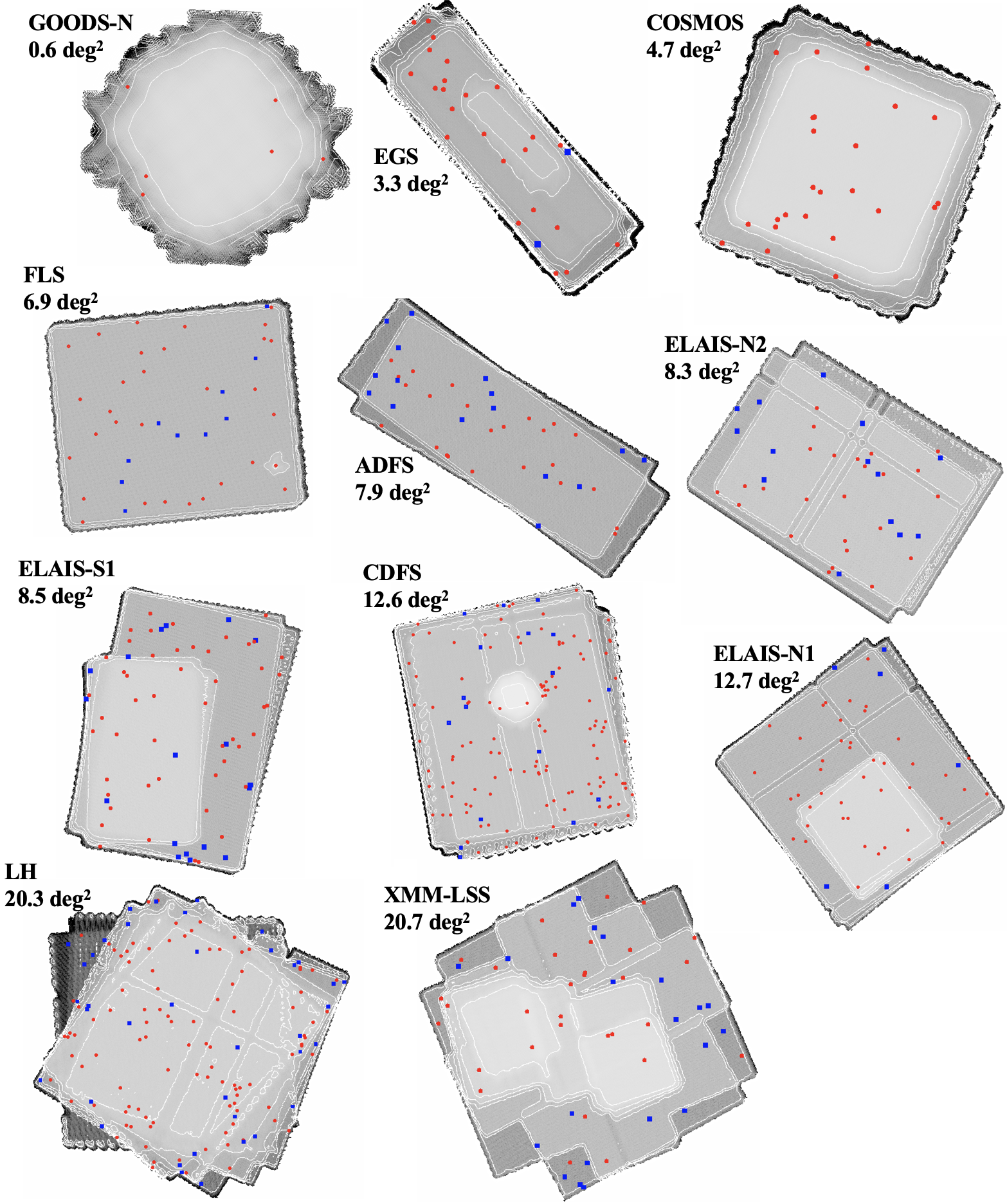}
\caption{Positions of the 500~$\mu$m risers in each field selected based on the
HerMES DR4 xID catalog (red circles) and the 500~$\mu$m residual map (blue
boxes), overlaid on top of the 250~$\mu$m error map displayed in inverted
gray scale. Smaller values on the error map (lighter regions) indicate deeper
exposures. White contours are plotted on the error maps for the illustration
purposes. The fields are shown in the order of their sizes (from left to
right, and from top to bottom), but the sizes are not displayed to scale.
}
\label{fig:500umRiserPos}
\end{figure*}

  In the rest of this paper, we will only discuss the Tier 1 objects. 
Figure~\ref{fig:500umRiserFluxColor} shows their flux density and color 
distributions. The median of $f_{500}$ for the xID-based and the residual-based
objects is 32.3 and 35.1~mJy, respectively, while that for the whole sample is
33.2~mJy. Figure~\ref{fig:500umRiserPos} shows their spatial distribution on 
top of the 250~$\mu$m error images that are used to indicate the sensitivity
levels of the survey in different areas (smaller errors, or lighter regions,
indicate higher sensitivities). 

\begin{longrotatetable}
\begin{deluxetable}{rcccccc|rcccccc}
\tablecolumns{14}
\tabletypesize{\scriptsize}
\tablecaption{Catalog of Tier 1 500~$\mu$m risers
    \label{tab:500RCatSample}
}
\tablehead{
\colhead{ID} &
\colhead{RA} &
\colhead{DEC} &
\colhead{$f_{250}$} & 
\colhead{$f_{350}$} & 
\colhead{$f_{500}$} & 
\colhead{Radio}  &
\colhead{ID} &
\colhead{RA} &
\colhead{DEC} &
\colhead{$f_{250}$} & 
\colhead{$f_{350}$} & 
\colhead{$f_{500}$} & 
\colhead{Radio}  \\
\colhead{} &
\colhead{(degrees)} &
\colhead{(degrees)} &
\colhead{(mJy)} &
\colhead{(mJy)} &
\colhead{(mJy)} &
\colhead{Code} &
\colhead{} &
\colhead{(degrees)} &
\colhead{(degrees)} &
\colhead{(mJy)} &
\colhead{(mJy)} &
\colhead{(mJy)} &
\colhead{Code} 
}
\startdata
500R\_CDFS\_T1\_x004 & 52.7319600 & $-$28.8153230 & $47.3\pm3.4$ & $59.0\pm3.9$ & $60.0\pm3.9$ &  1  & x176 & 52.5909230 & $-$27.7079180 & $17.7\pm3.4$ & $19.5\pm4.0$ & $27.2\pm3.9$ &  0 \\
                x007 & 52.5725400 & $-$27.5764300 & $37.5\pm3.4$ & $37.7\pm3.9$ & $45.4\pm3.9$ &  0  & x177 & 52.5599670 & $-$28.9544070 & $23.9\pm3.4$ & $25.6\pm4.0$ & $36.1\pm3.9$ & 99 \\
                x009 & 53.7353200 & $-$29.8309040 & $39.1\pm3.4$ & $47.1\pm3.9$ & $48.8\pm4.2$ & 99  & x178 & 53.5423740 & $-$28.6750700 & $16.3\pm3.4$ & $25.1\pm3.9$ & $30.2\pm4.0$ &  0 \\
                x162 & 52.6515000 & $-$26.7467480 & $18.1\pm3.4$ & $23.2\pm4.0$ & $23.9\pm4.0$ & 99  &  R32 & 53.9888420 & $-$27.9607340 & $ 9.6\pm3.4$ & $17.7\pm3.8$ & $22.0\pm4.0$ &  0 \\
                x166 & 51.7917800 & $-$28.3055460 & $19.5\pm3.4$ & $20.7\pm3.9$ & $22.3\pm4.2$ &  0  &  R33 & 54.0337730 & $-$27.8183910 & $11.6\pm3.4$ & $19.1\pm3.8$ & $21.7\pm3.9$ & 99 \\
                x170 & 52.7494700 & $-$28.8928720 & $18.2\pm3.4$ & $18.7\pm3.9$ & $20.9\pm4.1$ &  0  &  R37 & 52.8901080 & $-$26.8169130 & $ 9.5\pm3.4$ & $15.5\pm3.9$ & $20.6\pm3.9$ & 99 \\
\enddata
\tablecomments{Catalog of Tier 1 500~$\mu$m risers, using part of those
in the CDFS as examples. The entire catalog is available in Appendix.
When referred to in full, the object name consists of
the leading string and the sequential ID. The leading string has the field name
coded in. In addition, ``T1'' stands for ``Tier 1.'' For clarity, the leading 
string is omitted except in the first entry of the field. In the sequential ID, 
``x'' or ``R'' indicates the method with which the object is selected (using the 
HerMES xID catalog or the residual map, respectively). The equatorial coordinates
(J2000.0) are in degrees. The flux density and the total error are in the units of mJy. 
For those ``x'' objects, these are from the HerMES xID catalog, while for those ``R''
objects, these are based on our own photometry (see \S 3.2). The ``Radio Code''
indicates the status of the object in terms of radio observations
(see \S\ref{sec:500R_Radio} for details): ``1'' -- detected
in either FIRST and/or deeper radio data, ``0'' -- covered by deeper radio data but
not detected, and ``99'' -- outside of the coverage of deeper radio data. The radio
properties of those with code ``1'' and ''0'' are summarized in Tables
\ref{tab:500RtoFIRST}, \ref{tab:500RtoDeepRadio} and \ref{tab:500RnonRadio}.
}
\end{deluxetable}
\end{longrotatetable}

\section{Selection of SPIRE Dropouts as Potential 850~$\mu$m Risers}

   We further selected potential 850~$\mu$m risers in the five S2CLS fields.
It would be natural to expect that a subset of the 500~$\mu$m risers selected
above in these fields could be 850~$\mu$m risers. However, we found that none
of them are. This is in part due to the limited SPIRE survey sensitivities.
The noise in the HerMES maps is at the level of $\sim 4$--5 mJy (confusion 
noise included), which means that a HerMES source at S/N $\sim 5$ would have a
flux density of $\gtrsim 20$~mJy in the SPIRE bands. None of the S2CLS sources
in these fields have $S_{850}>20$~mJy.

   On the other hand, we noticed that some of the S2CLS sources are very weak
or invisible in the HerMES data. For simplicity, we shall refer to such
850~$\mu$m sources as the ``SPIRE dropouts.'' It is impossible to judge at this
stage whether these objects are 850~$\mu$m risers, again because the 
sensitivity of the SPIRE data is too shallow to sufficiently constrain the SEDs
at the bluer wavelengths. However, if 850~$\mu$m risers do exist in these S2CLS
fields, they must be among these SPIRE dropouts.

   We selected the SPIRE dropouts in a two-step approach. We only considered 
the most reliable 850~$\mu$m sources that have detection S/N
(``detection\_SNR'' in the S2CLS catalogs) $\geq 5$. 
We first cross-matched these S2CLS sources with the HerMES sources in the
xID catalogs, and found those that were missing from the latter. We refer to
this step as the selection using ``Method A.'' The matching
radius was chosen based on the positional uncertainties in
both. Following Equations 1 and 2 in \citet[][]{MY15} (see also
\citet[][]{Ivison2007b}),
the uncertainty $\mathrm{\sigma_{pos}}$ can be expressed as
\begin{equation}\label{eq:poserr}
  \mathrm{\sigma_{pos}} = \frac{0.6}{\text{S/N}}\sqrt{\theta_a^2 + \theta_b^2}
               = \frac{0.6\times 1.414\times \theta}{\text{S/N}},
\end{equation}
where $\theta_a$ and $\theta_b$ are the beam sizes along the major and
the minor axes, respectively, and $\theta$ is the total beam size when the
beam is symmetric. For the SCUBA2 850~$\mu$m and the SPIRE 250~$\mu$m, $\theta$
is 13\arcsec\, and 18\arcsec, respectively. We treated the 250$\mu$m 
positions as being reliable if the sources have $f_{250}/et_{250}\geq 3$. 
Plugging in S/N $=5$ and 3 for 850 and 250~$\mu$m, respectively, and adding in
quadrature the two terms, we obtained the positional matching error of 
$\mathrm{\sigma_{pos}}=5$.\arcsec55, which we adopted as the matching radius.

   In the next step, we also selected the matched 850~$\mu$m sources that have
S/N $<3$ in 250~$\mu$m. For a source below this limit, the SPIRE photometry is
prone to large errors; while the reported SPIRE flux densities are still 
brighter than the 850~$\mu$m one, the true situation could be the opposite.
Therefore, this step was to make certain that we would not miss the real SPIRE
dropouts whose ``detections'' in SPIRE are artificially boosted due to their
low S/N in the HerMES data. We refer to this secondary step as the selection
using ``Method B.''

   Similar to what was done with the 500~$\mu$m risers, all of the above
candidate SPIRE dropouts were visually examined to remove contaminations. The
most common contaminators are due to blending, especially in the SPIRE bands.
There are also cases where it was difficult to decide whether there was a
detection in any of the SPIRE bands, and if yes, whether the weak detection
would decrease the prospect of the candidate being a 850~$\mu$m riser. The
visual judgment on the latter was aided by using the reported SPIRE flux 
densities (despite them having large errors). Specifically, we required
$(S_{850}+0.5\times \mathrm{Err}S_{850})>(f_{250}-2\times et_{250})$ to keep
such a candidate, where $\mathrm{Err}S_{850}$ and $et_{250}$ are the errors in
850~$\mu$m and 250~$\mu$m, respectively. In both ``A'' and ``B'' categories, 
the most probable candidates were ranked as ``Tier 1,'' and the less probable
ones were ranked as ``Tier 2.'' In Figure 4, we show some examples in both 
tiers as well as some rejected contaminators. The full catalogs of the Tier 1 
objects are presented in Table \ref{tab:SD850CatSample}, 
while those of the Tier 2 objects are given in Appendix.
In the rest of this paper, we only discuss the Tier 1 objects.
Their flux density distribution is shown in Figure 5, while their spatial 
distributions in different fields are shown in Figure 6.

\begin{figure*}[t]
\plotone{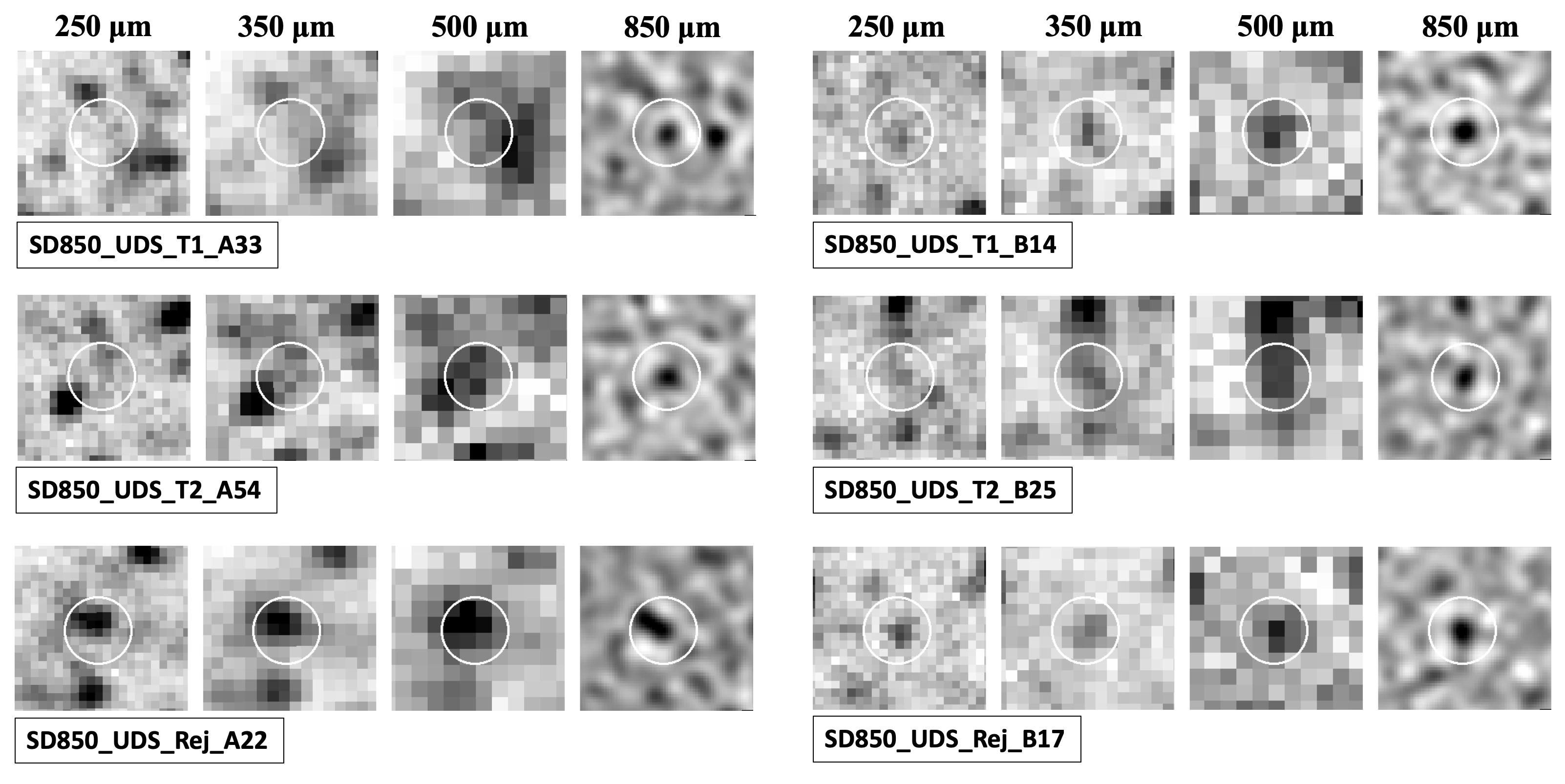}
\caption{Image stamps of six candidate SPIRE dropouts in the UDS field as 
examples. The objects have their IDs labeled in the black box, which all use
``SD850'' as the leading string. The ones shown on the left panel are selected
by Method ``A'' (designated by ``A'' in their IDs), which is to search for 
850~$\mu$m sources that have no counterparts in the HerMES 250~$\mu$m-based
xID catalog.
The ones shown in the right panel are selected by Method ``B'' (designated
by ``B'' in their IDs), which is to search for those that have very weak 
250~$\mu$m counterparts ($S/N<3$) in the xID catalog. The
stamps are 2\arcmin$\times$2\arcmin\, in size and are oriented North-up and
East-left. The white circles are 25\arcsec\, in radius and are centered on the
candidates. In each panel, (1) the images from left to right are the HerMES
250, 350, 500~$\mu$m and the S2CLS 850~$\mu$m, respectively, (2) the first row
is a Tier 1 object, designated by ``T1'' in the ID, (3) the second row is a
Tier 2 object, designated by ``T2'' in the ID, and (4) the last row is a 
rejected contaminator, designated by ``Rej'' in the ID. The contaminators 
shown here are representative of two common cases when the candidates have to
be rejected. The one in the left panel has a very close neighbor, which
severely contaminates the source position in 500~$\mu$m and therefore is
difficult to judge if it is a real SPIRE dropout. The one in the right panel,
despite being reported as a 250~$\mu$m source with $S/N<3$ in the xID catalog,
has very obvious detections in all three SPIRE bands, with reported
$f_{250}=13.6\pm 4.6$~mJy. Given its $S_{850}=5.6$~mJy, it is unlikely to be
brighter in 850~$\mu$m than in the SPIRE bands. 
}
\label{fig:SD850Demo}
\end{figure*}

\begin{figure}[]
\plotone{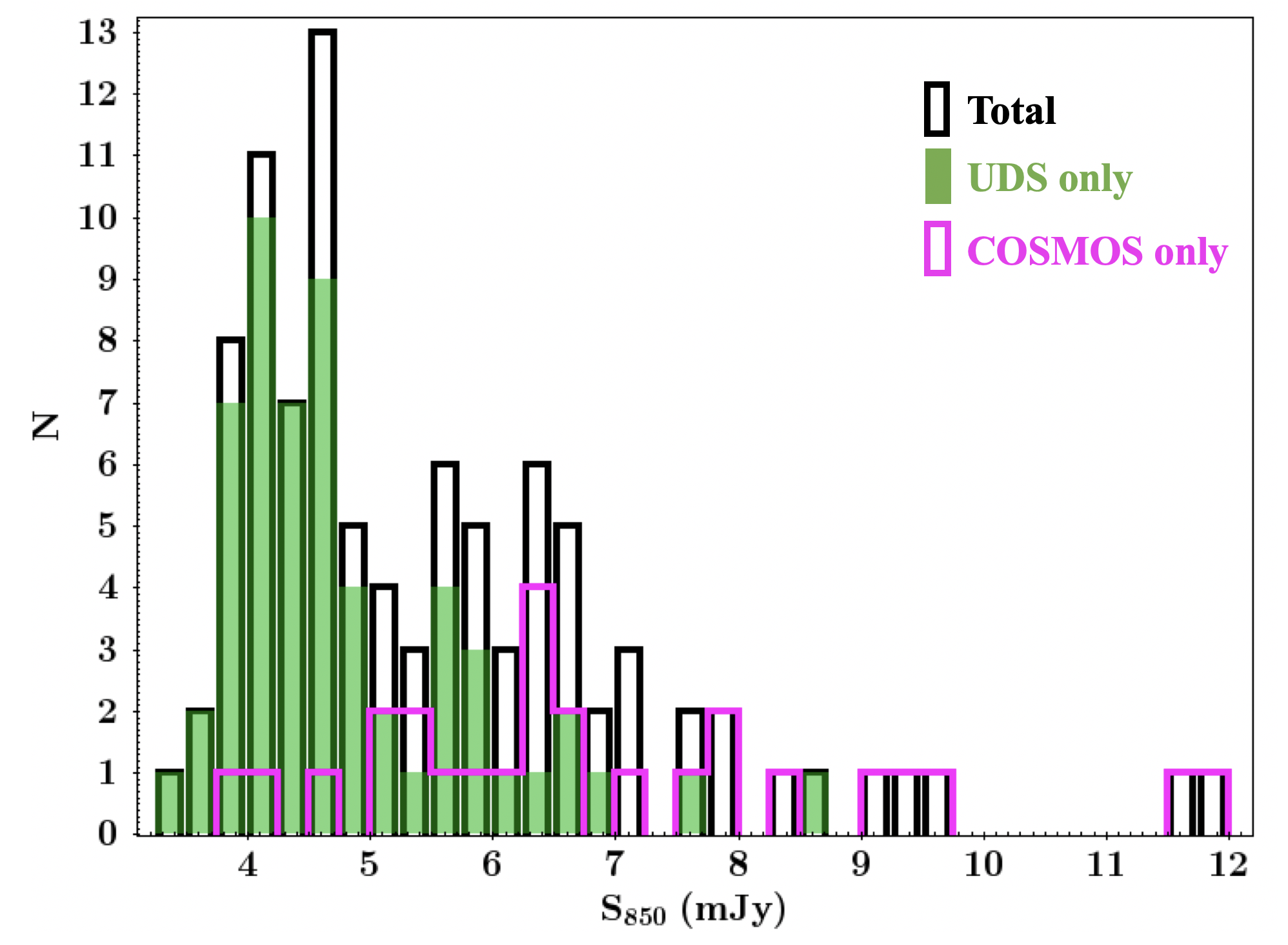}
\caption{850~$\mu$m flux density ($S_{850}$) distribution of the Tier 1 
SPIRE dropouts in our sample (95 objects in total). The objects selected in the
UDS field (57 objects; shown in green) make up 60.0\%, most of which are at
the faint end. Those in the COSMOS field (26 objects; shown in purple) are the
next largest group in the sample and make up 27.3\% of the total. There is a
clear offset of the representative $S_{850}$ values between the two subsets,
which is largely due to the different survey limits in these two fields.
}
\label{fig:SD850FluxHist}
\end{figure}

\begin{figure*}[t]
\plotone{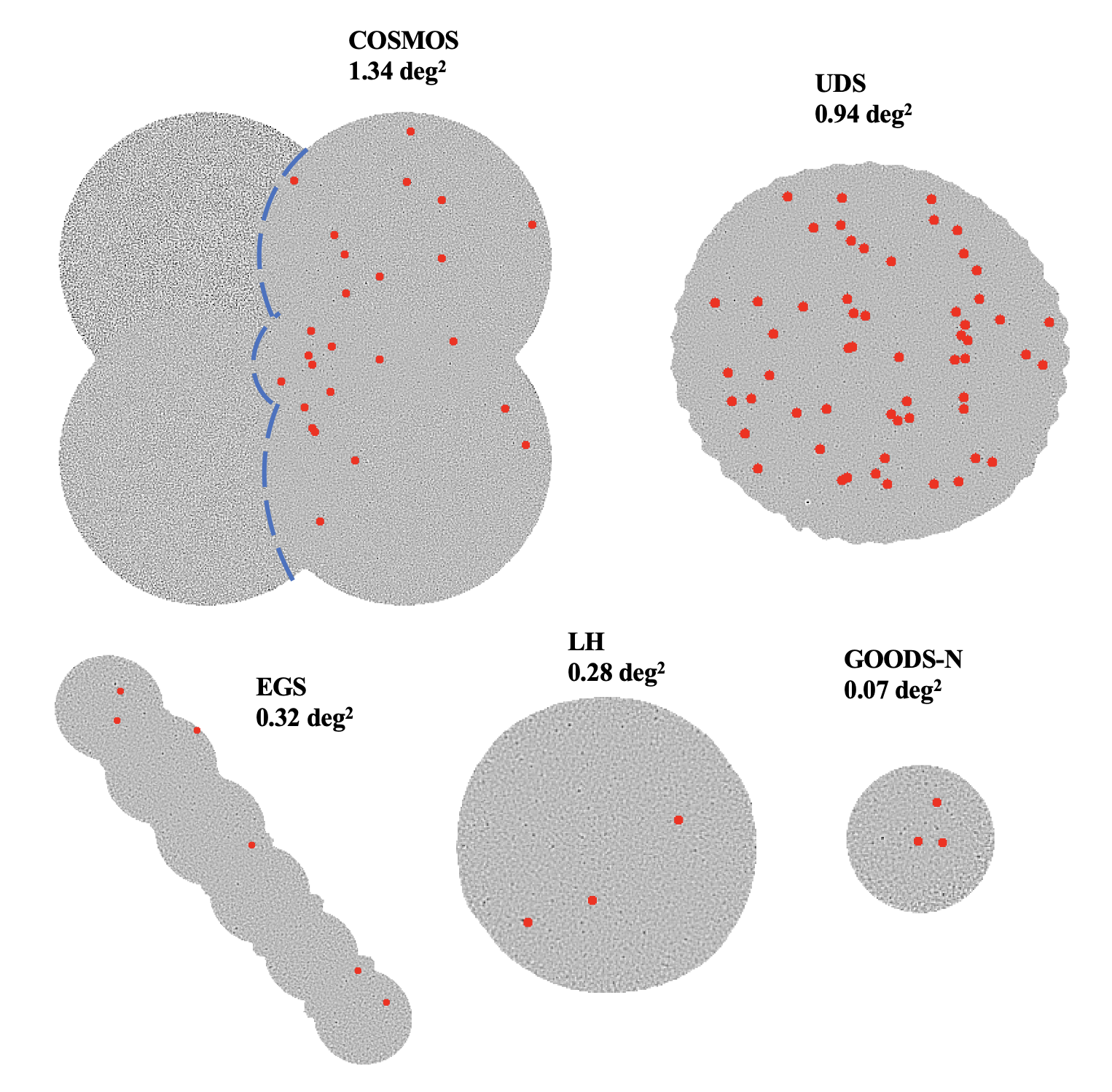}
\caption{Positions of the SPIRE dropouts in each field (red dots) selected
based on the S2CLS and the HerMES data, overlaid on the S2CLS 850~$\mu$m maps.
The fields are shown in decreasing order of their sizes (from left to
right, and from top to bottom), but the sizes are not displayed to scale.
Note that the released map of the COSMOS field include the data from the S2XLS,
however the S2CLS catalog (and hence our search in this field) is based on the
S2CLS data only (the western part separated by the dashed curve) and covered
1.34~deg$^2$.
}
\label{fig:SD850Pos}
\end{figure*}


\startlongtable
\begin{longrotatetable}
\begin{deluxetable}{rcccccccccccc}
\tablecolumns{12}
\tabletypesize{\scriptsize}
\tablecaption{Catalog of Tier 1 SPIRE dropouts based on S2CLS 850~$\mu$m data
   \label{tab:SD850CatSample}
}
\tablehead{
\colhead{Object Name} & 
\colhead{RA$\mathrm{_{850}}$} & 
\colhead{DEC$\mathrm{_{850}}$} &
\colhead{S/N$\mathrm{_{850}}$} &
\colhead{$S_{850}$} &
\colhead{RA$\mathrm{_{SPIRE}}$} &
\colhead{DEC$\mathrm{_{SPIRE}}$} & 
\colhead{$f_{250}$} & 
\colhead{$f_{350}$} & 
\colhead{$f_{500}$} &
\colhead{$\Delta$} &
\colhead{Radio} \\
\colhead{} &
\colhead{(degrees)} &
\colhead{(degrees)} &
\colhead{} &
\colhead{(mJy)} &
\colhead{(degrees)} &
\colhead{(degrees)} &
\colhead{(mJy)} &
\colhead{(mJy)} &
\colhead{(mJy)} &
\colhead{(\arcsec)} &
\colhead{Code}
}
\startdata
SD850\_COSMOS\_T1\_A02 & 150.038273 &    2.140164 & 9.0 & $9.0\pm1.3$ &    \nodata   &    \nodata &      \nodata  &      \nodata  &      \nodata  & \nodata & 0 \\
                   A03 & 149.989321 &    2.457937 & 8.7 & $11.8\pm1.9$ &   \nodata   &    \nodata &      \nodata  &      \nodata  &      \nodata  & \nodata & 0 \\
                   A04 & 149.792953 &    2.815101 & 8.3 & $11.7\pm2.1$ &   \nodata   &    \nodata &      \nodata  &      \nodata  &      \nodata  & \nodata & 1 \\
                   B03 & 150.109433 &    2.257389 & 6.7 & $5.5\pm1.1$ & 150.109220 &   2.256556 & $ 2.5\pm3.4$ & $ 3.4\pm6.6$ & $ 0.0\pm13.0$ & 3.1 & 0 \\
                   B05 & 150.027677 &    2.642935 & 6.3 & $8.0\pm1.8$ & 150.027700 &   2.642787 & $ 3.9\pm3.4$ & $ 8.5\pm6.6$ & $12.9\pm5.0$ & 0.5 & 0 \\
                   B06 & 150.088867 &    2.010723 & 6.0 & $6.3\pm1.4$ & 150.087950 &   2.011739 & $ 3.7\pm3.4$ & $ 5.5\pm6.6$ & $ 5.2\pm5.2$ & 4.9 & 1\\
\enddata
\tablecomments{Catalog of Tier 1 SPIRE dropouts, using part of the entries 
in the COSMOS field as examples. The entire catalog is available in Appendix.
When referred to in full, the object name consists of
the leading string and the sequential ID. The leading string has the field name
coded in. In addition, ``T1'' stands for ``Tier 1''. For clarity, the leading 
string is omitted except in the first entry of the field. In the sequential ID, 
``A'' or ``B'' indicates that the object is selected by Method A or B (see \S 4).
$\mathrm{RA_{850}}$ and $\mathrm{DEC_{850}}$ are the equatorial coordinates (J2000.0)
based on the positions given in the S2CLS catalog. The ``deboosted'' flux density
in 850~$\mu$m from that catalog is reported under $S_{850}$. The ``B''
objects have weak SPIRE counterparts in the HerMES xID catalog, and their 
SPIRE positions and flux densities are listed under $\mathrm{RA_{SPIRE}}$ and
$\mathrm{DEC_{SPIRE}}$, $f_{250}$, $f_{350}$, and $f_{500}$, respectively.
The seperation between their SCUBA2 and SPIRE positions is given under ``$\Delta$''.
``Radio Code'' indicates whether the object is detected in the radio data (``1'')
or not (``0''), or is outside of the radio data coverage (``99''), which is
detailed in \S\ref{sec:SD850_Radio}. The radio properties of those with code
``1'' and ``0'' are summarized in Table \ref{tab:SD850toDeepRadio} and
\ref{tab:SD850nonRadio}, respectively.
}
\end{deluxetable}
\end{longrotatetable}

\section{Comparison to the literature}

   While limited, there are some similar, very red FIR/sub-mm galaxies in
the literature that are found in regions in common with this work. 
Here we compare these to our samples.

\subsection{Comparison of 500~$\mu$m risers}

   The largest sample that we compare to is that of \citet[][]{Dowell2014},
which consists of 1, 18 and 21 500~$\mu$m risers in the GOODS-N, the FLS, and
the LH fields, respectively. 

   The one in the GOODS-N, their GOODSN 8, was selected as a candidate from the
xID catalog in our first step; however, it was removed from the final catalog
due to a blending problem.

   In the FLS field, 8 of their 18 objects are in our Tier 1 catalog, namely
their FLS 3, 5, 17, 19, 24, 25, 30, and 33 (FLS 3 is the $z=6.34$ object HFLS3
reported in \citet[][]{Riechers2013}). In addition, their FLS 20 is in our
Tier 2 catalog. Five other objects are not in our sample but can be accounted
for, namely, FLS 1, 6, 26, 31, and 32 (FLS 1 is at $z=4.29$ as reported in 
\citet[][]{Dowell2014}). Based on the DR4 xID catalog, the first four of them
have $f_{350}>f_{500}$ and $f_{350}>f_{250}$, and thus are 
``350~$\mu$m peakers'' instead of 500~$\mu$m risers. FLS 32 has S/N $<5$ in
500~$\mu$m and hence does not meet our criteria. The remaining four objects,
FLS 7, 22, 23, and 34, are not included either in the xID catalog or the
catalog generated based on the photometry in the 500~$\mu$m residual map. Upon
visual examination of the images, we find that (1) FLS 7 is severely affected
by blending due to a much brighter neighbor; (2) FLS 22 is too close to the
field edge, and neither the xID nor our own extraction on the residual map
includes this object; (3) FLS 23 is missed most likely due to its faintness;
and (4) FLS 34 must be a misidentification due to blending and thus not a
real source.

   In the LH field, three of their sources are included in our Tier 1 catalog,
namely, their LSW 20, 25, and 102 (LSW 20 is at $z=3.36$ as reported in
\citet[][]{Dowell2014}). Their LSW 28 was selected as a candidate in our
first step but was later rejected due to the severe blending problem and its
being too weak. Twelve of their sources are in the xID catalog (their
LSW 26, 29, 31, 47, 48, 49, 50, 52, 53, 54, 55, 56, 58, 60, 76, 81, and 82) but
were not selected by our procedure because they are all 350~$\mu$m peakers 
instead of 500~$\mu$m risers. The rest five objects, their LSW 49, 50, 55, 56,
and 82 are not in either the xID catalog nor our catalog based on the residual
map. Among them, LSW 50 is too faint in both 250 and 500~$\mu$m to be
included, and the other four all suffer from severe blending.

   The ``prototype'' 870~$\mu$m riser in \citet[][]{Riechers2017}, their 
ADFS-27, is a 500~$\mu$m riser in the first place (but not a SPIRE dropout);
however, it is not included in our sample because it has $\mathrm{S/N}=4.4<5$ in
500~$\mu$m.

   We note that two famous SMGs with known redshifts in GOODS-N, GN10 at 
$z=4.04$ and GN20 at $z=4.05$ \citep[][]{Daddi2009a, Daddi2009b}, are also not
included in our sample. The former one is severely blended with neighbors and
thus is not extracted, while the latter is actually a 350~$\mu$m peaker based on
the xID catalog.

\subsection{Comparison of SPIRE Dropouts}

   The only known SPIRE dropouts that we can compare to are the two millimeter
galaxies reported by \citet[][]{Ikarashi2017} in the XMM-LSS field, which were
first detected by using ASTE/AzTEC at 1.1 mm \cite[][]{Ikarashi2015}. Based on
the subsequent identifications in the HerMES and the S2CLS data, 
\citet[][]{Ikarashi2017} point out that they are secure detections in 
850~$\mu$m but are invisible in the SPIRE data. These two objects, their 
ASXDF1100.053.1 and 231.1, are recovered in our SPIRE dropouts and correspond
to our \texttt{SD850\_UDS\_T1\_A36} and \texttt{A42}, respectively.

\section{Radio Properties}

   It is well known that FIR galaxies dominated by dust-embedded star 
formation follow a tight FIR-radio relation \citep[see][for review]{Condon1992}.
The corollary is that their observed spectral indices between FIR/sub-mm and
radio wavelengths should be a strong function of redshift due to the opposing
K-corrections in these two regimes \citep[see e.g.][]{CarilliYun1999}. As a
result, high-redshift dusty star forming galaxies in general should be 
radio weak unless they also host an active galactic nucleus (AGN). Therefore, 
we examine the radio properties of our candidates whenever sufficiently deep 
radio data are available. We first examine them against the public data from
the Faint Images of the Radio Sky at Twenty-cm 
\citep[FIRST;][]{Becker1995}, which covered over 10,000 deg$^2$ of the Galactic
Caps at 1.4~GHz to the typical sensitivity of rms $\sim 0.15$~mJy~beam$^{-1}$.
Some of our fields have deeper radio data, which we subsequently examine
as well. While in most cases they only cover a fraction of the area, these are
valuable in further selecting the most promising high-redshift candidates.

\subsection{Radio Properties of 500~$\mu$m risers\label{sec:500R_Radio}}

   In determining the matching radius to the radio sources, we consider the
positional errors due to both the SPIRE and the radio beams
($\mathrm{\sigma_{pos}^{SPIRE}}$ and $\mathrm{\sigma_{pos}^{radio}}$, 
respectively), which are calculated based on Equation \ref{eq:poserr} and then
are added in quadrature to obtain the total error 
$\mathrm{\sigma_{pos}}=\sqrt{(\mathrm{\sigma_{pos}^{SPIRE}})^2+(\mathrm{\sigma_{pos}^{radio}})^2}$.
In most cases, the former is the dominant term. For the xID-based and the 
residual-based objects, we use $\theta=18$\arcsec\ and 36\arcsec\ in Equation
\ref{eq:poserr}, which are the 250 and the 500 $\mu$m beam sizes, respectively.
We take this approach because the positions of these two sets are based on the
250 and the 500~$\mu$m maps, respectively. For reference, these correspond to
$\mathrm{\sigma_{pos}^{SPIRE}}=3$.\arcsec 05 and 6.\arcsec 11, respectively, at
$\mathrm{S/N=5}$. For different objects, $\mathrm{\sigma_{pos}^{SPIRE}}$
vary because of their different S/Ns in SPIRE.
As the radio positional error is usually much smaller, we do not consider the
radio source S/N individually but calculate at the fixed $\mathrm{S/N=5}$.
To simplify the matching process, we search for the radio counterparts within a
generous matching radius of $r=10$\arcsec, calculate the positional offsets,
$\mathrm{\Delta_{pos}}$, between our objects and the matched radio sources,
and determine the ratios $\Delta/\sigma$ between $\mathrm{\Delta_{pos}}$ and
$\mathrm{\sigma_{pos}}$. The matches with $\Delta/\sigma\leq 1.5$ are deemed
to be reliable and are listed in Table \ref{tab:500RtoFIRST} and 
\ref{tab:500RtoDeepRadio}. The unmatched ones are indicated
as such in Table \ref{tab:500RCat}, and we quote their upper limits based on
the relevant radio catalogs. We discuss both the matched and the unmatched 
sources below in some detail.


\subsubsection{Cross-matching with the FIRST sources}

  All of our equatorial and northern fields are covered by FIRST, which, albeit
being shallow, is useful in detecting the most obvious radio AGNs. The
angular resolution of FIRST is 5\arcsec, and therefore 
$\mathrm{\sigma_{pos}^{radio}}=$~0.\arcsec 85.

  As it turns out, none of our objects in COSMOS, EGS, ELAISN1, and GOODS-N have
FIRST matches within 10\arcsec. In the other four equatorial and northern
fields, we have seven matches with $\Delta/\sigma\leq 1.5$, which are summarized
in the top part of Table \ref{tab:500RtoFIRST}.
Four of these radio counterparts have monotonically increasing
SED from FIR to radio, suggesting that they are likely blazars
\citep[see][Section 4.7]{MY15}. \texttt{500R\_ELAISN2\_T1\_x03} and
\texttt{500R\_LH\_T1\_x003} are indeed listed as blazars in the Fermi Large
Area Telescope AGN catalog \citep[][]{Ackermann2015}. The former one is
associated with SDSS J163915.80+412833.7 at $z=0.69$ \citep[][]{Schneider2010}. 
\texttt{500R\_FLS\_T1\_x07} is also listed as a blazar by 
\citet[][]{Marcha2013} and \citet[][]{Mingaliev2014}, and has the known 
redshift of $z=0.2974$ \citep[][]{Marleau2007}. Finally, 
\texttt{500R\_LH\_T1\_x039} is also likely a blazar and has $z=0.5795$ based 
on the SDSS DR13. Therefore, the contamination of the 500~$\mu$m riser sample
due to blazars is only $\sim 1.0$\% (4 out of the 381 total in the northern
and equatorial fields). The other three objects have $S_{1.4}>1$~mJy and must
host strong AGNs.

   Table \ref{tab:500RtoFIRST} summarizes the above results.

\begin{table*}
\scriptsize
\centering
\caption{Match of Tier 1 500~$\mu$m risers to FIRST
    \label{tab:500RtoFIRST}}
\begin{ctabular}{lcccccccl}
\toprule
Object Name & $\mathrm{\sigma_{pos}}$ & $\mathrm{\Delta_{pos}}$ & 
$\Delta$/$\sigma$ & RA$_F$ & DEC$_F$ & $S_{1.4}$ & $\mathrm{\sigma_{1.4}}$ &
Comments \\
            & \multicolumn{1}{c}{(\arcsec)} & \multicolumn{1}{c}{(\arcsec)} &
            &             
            &
            & \multicolumn{1}{c}{(mJy)}     
            & \multicolumn{1}{c}{(mJy~beam$^{-1}$)}    &          \\
\midrule
 500R\_ELAISN2\_T1\_x03 & 1.9 & 2.6 & 1.4 & 16:39:15.813 & $+$41:28:33.61 &  91.76 & 0.161 & blazar\tablenotemark{a}; $z=0.69$\tablenotemark{b}\\
 500R\_ELAISN2\_T1\_x49 & 3.5 & 1.7 & 0.5 & 16:36:46.387 & $+$40:14:36.67 &   7.25 & 0.143 & ...\\
 500R\_FLS\_T1\_x07     & 2.4 & 0.8 & 0.3 & 17:25:35.004 & $+$58:51:39.92 &  71.06 & 0.147 & blazar\tablenotemark{c,}\tablenotemark{d}; $z=0.2874$\tablenotemark{e}\\
 500R\_FLS\_T1\_x50     & 4.4 & 6.0 & 1.4 & 17:23:15.537 & $+$59:09:18.93 &   1.13 & 0.143 & ...\\
 500R\_LH\_T1\_x003     & 1.4 & 0.6 & 0.4 & 10:37:44.322 & $+$57:11:55.57 & 130.02 & 0.147 & blazar\tablenotemark{a}\\
 500R\_LH\_T1\_x039     & 2.3 & 1.4 & 0.6 & 10:40:37.806 & $+$55:40:52.73 &  61.39 & 0.144 & likely blazar; $z=0.5795$\tablenotemark{f}\\
 500R\_XMMLSS\_T1\_x83  & 4.7 & 4.6 & 1.0 & 02:19:19.156 & $-$06:45:54.16 &   1.77 & 0.143 & ...\\
\hline
 500R\_FLS\_T1\_x36     & 3.4 & 9.1 & 2.7 & 17:06:42.635 & $+$60:37:57.43 &   3.59 & 0.331 & ...\\
 500R\_FLS\_T1\_x03     & 2.4 & 7.5 & 3.1 & 17:08:52.043 & $+$58:48:56.87 &   1.67 & 0.137 & ...\\
 500R\_FLS\_T1\_R11     & 5.2 & 9.5 & 1.8 & 17:11:31.799 & $+$59:39:24.99 &  14.22 & 0.149 & ...\\
 500R\_LH\_T1\_R06      & 4.0 & 8.8 & 2.2 & 11:00:48.619 & $+$59:20:09.32 &   9.49 & 0.155 & ...\\
\bottomrule
\end{ctabular}
\tablecomments{Tier 1 500~$\mu$m risers matched to the radio sources in
the FIRST catalog. The matching radius is $r=10$\arcsec.
$\mathrm{\sigma_{pos}}$ is the total positional error due to the uncertainties
of the SPIRE and the FIRST data. $\mathrm{\Delta_{pos}}$ is the
separation between the SPIRE and the FIRST positions. $\Delta/\sigma$ is the
ratio of the two, and only the top seven matches with $\Delta/\sigma\leq 1.5$
are deemed to be reliable. The bottom four are listed for reference. The
equatorial coordinates (J2000.0) are the FIRST positions. $S_{1.4}$ and 
$\sigma_{1.4}$ are the integrated flux density and the RMS sensitivity taken 
from the FIRST catalog, respectively. The references for the blazar 
identifications and the redshifts are (a)\citet[][]{Ackermann2015},
(b) \citet[][]{Schneider2010}, (c) \citet[][]{Marcha2013}, 
(d) \citet[][]{Mingaliev2014}, (e) \citet[][]{Marleau2007}, and
(f) SDSS DR13.
}
\end{table*}

\begin{table*}
\centering
\caption{Summary of matching Tier 1 500~$\mu$m risers to deep radio data
    \label{tab:500RDeepRadioSummary}}
\begin{tabular}{llcccccc}
\toprule
Field & Facility & Area    & Beam Size & $\mathrm{\sigma_{pos}^{Radio}}$ & Sensitivity & Obj Coverage & Obj Matched \\
      &          & \multicolumn{1}{c}{(deg$^2$)} & \multicolumn{1}{c}{(\arcsec)} & \multicolumn{1}{c}{(\arcsec)} & \multicolumn{1}{c}{($\mu$Jy)} & \\
\midrule
 ADFS      & ATCA, 1.4~GHz & 2.5   & 6.2$\times$4.9 & 0.95  & 18--200 & 10/40  &  0 \\
 CDFS      & VLA, 1.4~GHz  & 0.324 & 2.8$\times$1.6 & 0.39  & 37.0    &  2/140 &  0 \\
           & AT, 1.4~GHz   & 3.6   & 16$\times$7    & 2.10  & 70      & 32/140 &  8 \\
 COSMOS    & VLA, 3.0~GHz  & 2     & 0.75           & 0.13  & 11.5    & 17/25  & 10 \\
 EGS       & VLA, 1.4~GHz  & 0.73  & 3.8            & 0.66  & 130     &  8/26  &  2 \\
 ELAISN1   & GMRT, 610~MHz & 9     & 6$\times$5     & 0.94  & 350     & 35/51  &  0 \\
 ELAISN2   & GMRT, 610~MHz & 6     & 6.5$\times$5   & 0.98  & 425     & 31/39  &  2 \\
 ELAISS1   & AT, 1.4~GHz   & 2.7   & 16$\times$7    & 2.10  &  75     & 16/68  &  0 \\
 FLS       & VLA, 1.4~GHz  & 5     & 5              & 0.86  & 115     & 27/41  &  6 \\
 GOODS-N   & VLA, 1.4~GHz  & 0.44  & 1.7            & 0.29  &  20     &  4/6   &  1 \\
 LH        & GMRT, 610~MHz & 13    & 6$\times$5     & 0.94  & 300     & 89/141 &  3 \\
 XMM-LSS   & VLA, 1.4~GHz  & 0.8   & 5$\times$4     & 0.77  & 100     &  1/52  &  0 \\
\bottomrule
\end{tabular}
\tablecomments{The values of $\mathrm{\sigma_{pos}^{Radio}}$ are calculated based on 
Equation \ref{eq:poserr} and $\mathrm{S/N=5}$. The sensitivity levels are
based on the source catalogs that we used, and in most cases correspond to
$\mathrm{S/N=5}$ (see \S 6.1.2 for details). For the data that are not obtained 
at 1.4~GHz, the conversion to 1.4~GHz for the quoted sensitivity levels can
done by assuming a power law SED of $S_{\nu} \sim \nu^{-0.8}$.
The number pairs under ``Obj Coverage'' are those of the objects in the areas
covered by the relevant radio data and the total in the Tier 1 samples, 
respectively, while the numbers under `` Obj Matched'' are those of the objects
detected above the quoted sensitivity levels.
}
\end{table*}

\begin{table*}
\scriptsize
\centering
\caption{Match of Tier 1 500~$\mu$m risers to deep radio data
    \label{tab:500RtoDeepRadio}}
\begin{tabular}{lccccccccl}
\toprule
Object Name & $\mathrm{\sigma_{pos}}$ & $\mathrm{\Delta_{pos}}$ &
$\Delta$/$\sigma$ & $f_{500}$ & RA$\mathrm{_{radio}}$ & DEC$\mathrm{_{radio}}$ & $S\mathrm{_{radio}}$ &
Hi$z$Idx & References \\
            & \multicolumn{1}{c}{(\arcsec)} & \multicolumn{1}{c}{(\arcsec)} &
            & \multicolumn{1}{c}{(mJy)}
            &
            &
            & \multicolumn{1}{c}{($\mu$Jy)}
            & \multicolumn{1}{c}{(500)}  &          \\
\midrule
   500R\_CDFS\_T1\_x004 & 2.4 & 3.1 & 1.3  & $60.0\pm3.9$ & 3:30:55.67 & $-$28:48:55.16 & $10863\pm558$ & 0.01 & (a)  \\
   500R\_CDFS\_T1\_x023 & 2.8 & 1.4 & 0.5  & $41.1\pm3.9$ & 3:30:25.84 & $-$27:38:10.41 &  $161\pm25$ & 0.26 & (a) \\
   500R\_CDFS\_T1\_x104 & 3.5 & 4.7 & 1.3  & $26.0\pm3.9$ & 3:33:35.54 & $-$28:41:32.72 &  $286\pm22$ & 0.09 & (a) \\
   500R\_CDFS\_T1\_x107 & 3.0 & 1.2 & 0.4  & $36.0\pm3.9$ & 3:29:35.25 & $-$27:39:23.39 &  $106\pm18$ & 0.34 & (a) \\
   500R\_CDFS\_T1\_x183 & 3.9 & 4.7 & 1.2  & $26.0\pm3.9$ & 3:26:37.13 & $-$28:46:11.38 &  $945\pm61$ & 0.03 & (a) \\
   500R\_CDFS\_T1\_x187 & 3.7 & 2.4 & 0.6  & $22.0\pm4.0$ & 3:30:34.00 & $-$27:39:43.26 &  $129\pm24$ & 0.17 & (a) \\
   500R\_CDFS\_T1\_x193 & 3.4 & 2.7 & 0.8  & $21.9\pm3.9$ & 3:26:44.36 & $-$28:10:52.01 &  $159\pm24$ & 0.14 & (a) \\
   500R\_CDFS\_T1\_x282 & 5.0 & 1.4 & 0.3  & $24.7\pm3.9$ & 3:29:22.71 & $-$27:21:53.06 &  $548\pm41$ & 0.05 & (a) \\
  500R\_COSMOS\_T1\_x01 & 2.9 & 2.5 & 0.9  & $26.2\pm5.1$ &  9:59:21.57 &   1:47:37.45 &   $25.5\pm2.6$ & 0.56 & (b)  \\
  500R\_COSMOS\_T1\_x05 & 2.2 & 1.6 & 0.7  & $27.4\pm5.0$ & 10:01:38.55 &   2:37:36.71 &   $29.1\pm2.8$ & 0.51 & (b)  \\
  500R\_COSMOS\_T1\_x07 & 2.5 & 2.9 & 1.2  & $27.9\pm5.2$ & 10:03:01.79 &   1:39:10.52 &   $35.8\pm3.5$ & 0.42 & (b)  \\
  500R\_COSMOS\_T1\_x08 & 2.2 & 0.3 & 0.1  & $28.1\pm5.0$ & 10:01:42.20 &   2:37:27.10 &   $18.7\pm2.4$ & $\mathbf{0.82^{*}}$ & (b)  \\
  500R\_COSMOS\_T1\_x13 & 2.2 & 1.2 & 0.5  & $28.6\pm5.1$ &  9:58:45.94 &   2:43:29.27 &   $56.1\pm3.6$ & 0.28 & (b)  \\
  500R\_COSMOS\_T1\_x23 & 2.1 & 2.5 & 1.2  & $31.1\pm5.2$ & 10:01:57.69 &   1:44:46.97 &   $89.8\pm5.0$ & 0.19 & (b)  \\
  500R\_COSMOS\_T1\_x24 & 2.7 & 4.0 & 1.5  & $31.4\pm5.2$ & 10:02:40.43 &   1:45:44.11 &   $12.7\pm2.4$ & $\mathbf{1.34^{*}}$ & (b)  \\
  500R\_COSMOS\_T1\_x25 & 3.8 & 2.5 & 0.7  & $31.4\pm5.3$ & 10:00:17.35 &   1:58:27.82 &   $46.4\pm3.2$ & 0.37 & (b)  \\
  500R\_COSMOS\_T1\_x26 & 3.9 & 4.0 & 1.0  & $32.2\pm5.1$ & 10:00:59.18 &   1:33:06.73 &   $24.3\pm3.8$ & $\mathbf{0.72^{*}}$ & (b)  \\
  500R\_COSMOS\_T1\_x31 & 2.6 & 1.5 & 0.6  & $36.0\pm5.2$ & 10:01:26.02 &   1:57:51.32 &   $24.1\pm2.6$ & $\mathbf{0.81^{*}}$ & (b)  \\
     500R\_EGS\_T1\_x10 & 2.1 & 0.8 & 0.4  & $44.0\pm4.3$ & 14:22:33.33 &  53:14:18.95 &   $67\pm12$ & 0.65 & (c) \\
     500R\_EGS\_T1\_x27 & 3.6 & 1.9 & 0.5  & $22.0\pm4.4$ & 14:24:01.70 &  53:23:19.47 &  $611\pm13$ & 0.04 & (c) \\
     500R\_FLS\_T1\_x11 & 3.1 & 1.8 & 0.6  & $35.0\pm5.8$ & 17:18:51.84 &  59:54:03.84 &  $148\pm25$ & 0.24 & (d) \\
     500R\_FLS\_T1\_x28 & 3.2 & 3.3 & 1.0  & $34.7\pm6.9$ & 17:14:40.46 &  58:20:35.02 &  $135\pm26$ & 0.26 & (d) \\
     500R\_FLS\_T1\_x30 & 3.6 & 3.8 & 1.1  & $41.6\pm5.8$ & 17:18:39.95 &  58:20:37.13 &  $129\pm22$ & 0.32 & (d) \\
  500R\_GOODSN\_T1\_x20 & 3.0 & 4.4 & 1.5  & $18.9\pm3.4$ & 12:35:09.43 &  62:11:13.40 &   $41.4\pm7.7$ & 0.46 & (e) \\
     500R\_LH\_T1\_x204 & 3.1 & 0.8 & 0.3  & $37.1\pm4.7$ & 10:50:31.14 &  56:30:45.9  & $1134\pm121$ & 0.03 & (f) \\
     500R\_LH\_T1\_x262 & 4.6 & 3.2 & 0.7  & $27.5\pm4.5$ & 10:42:50.88 &  57:31:19.7  &  $518\pm78$ & 0.05 & (f) \\
\hline
    500R\_ADFS\_T1\_x14 & 2.9 & 4.8 & 1.7  & $29.4\pm5.6$ & 4:46:58.70 & $-$53:23:29.90 &  $304\pm38$ & \nodata & (g)  \\
  500R\_COSMOS\_T1\_x02 & 2.8 & 6.1 & 2.2  & $26.4\pm5.0$ & 10:01:42.55 &   2:00:14.69 &  $109.0\pm6.3$ & \nodata & (c)  \\
  500R\_COSMOS\_T1\_x10 & 2.2 & 4.2 & 1.9  & $28.3\pm5.2$ & 10:01:40.44 &   2:30:10.44 &   $11.6\pm2.3$ & \nodata & (c)  \\
  500R\_COSMOS\_T1\_x34 & 1.7 & 6.0 & 3.5  & $39.8\pm5.1$ & 10:00:09.49 &   2:22:19.48 &  $142.0\pm7.5$ & \nodata & (c)  \\
  500R\_GOODSN\_T1\_x01 & 1.1 & 1.9 & 1.7  & $43.5\pm3.5$ & 12:39:05.94 &  62:05:36.40 &   $72.9\pm14.0$ & \nodata & (e)  \\
\bottomrule
\end{tabular}
\tablecomments{Tier 1 500~$\mu$m risers matched to the radio sources in various
surveys that are deeper than FIRST. The matching radius is $r=10$\arcsec. Only
the top 26 matches that have $\Delta/\sigma\leq 1.5$ are deemed to be reliable.
``Hi$z$Idx(500)'' is calculated based on Equation \ref{eq:highzindex}. Those
that with Hi$z$Idx(500) $\geq 0.7$ (potentially at $z>6$; see \S 6.4.1) are
boldfaced and marked with asterisk. The bottom five objects only have lower
limits of Hi$z$Idx (reported in Table \ref{tab:500RnonRadio}) because they are 
treated as having no radio counterparts and thus only have upper limits of 
their radio flux densities. The radio catalogs are from the references listed
in the last column. These are (a) \citet[][]{Franzen2015}, 1.4~GHz; 
(b) \citet[][]{Smolcic2017}, 3~GHz; (c) \citet[][]{Ivison2007a}, 1.4~GHz; 
(d) \citet[][]{Condon2003}, 1.4~GHz; (e) \citet[][]{Morrison2010}, 1.4~GHz; 
(f) \citet[][]{Garn2008b,Garn2010}, 610~MHz; (g) \citet[][]{White2012}, 1.4~GHz.
}
\end{table*}

\subsubsection{Cross-matching with deeper radio data\label{sec:SD850_Radio}}

   All our fields have partial radio data deeper than FIRST, which we further
use for the radio counterpart identification. Unfortunately, most of the radio
images are not publicly available, and hence our discussion below is based on
the catalogs published by the respective authors. 
Table \ref{tab:500RDeepRadioSummary} summarizes the matching results, and the
details are provided below for each field.

   --- {\it ADFS}\,\,\, Currently, the only deep radio data in this field are
those of \citet[][]{White2012}, which were obtained at 1.4~GHz using the
Australia Telescope Compact Array (ATCA). The beam size is 
6.\arcsec 2$\times$4.\arcsec 9, which corresponds to 
$\mathrm{\sigma_{pos}^{ATCA}}=$~0.\arcsec 95.
Ten of our objects are within the coverage, and only one 
(\texttt{500R\_ADFS\_T1\_x14}) is matched within 10\arcsec. However,
it has $\Delta/\sigma=1.7$ and is not regarded as a real match.

    The sensitivity of this radio map varies greatly from the field center
(1~$\sigma$ sensitivity, or rms, of 18-50~$\mu$Jy~beam$^{-1}$) to the edge
($\sim 200$~$\mu$Jy~beam$^{-1}$). We cannot obtain the brightness upper limit
that is close to the real sensitivity limit at the location of each
nondetection because we do not have the map at hand. We choose to assign a
universal, conservative upper limit of 1~mJy to all these nondetections,
which corresponds to $\sim 5 \sigma$ noise at the field edge.

   --- {\it CDFS}\,\,\, There have been two 1.4~GHz surveys in this field, 
which are described in \citet[][]{Miller2013} and \citet[][]{Franzen2015},
respectively. The data from the former were obtained at the Very Large
Array (VLA) and have reached the rms of 7.4~$\mu$Jy~beam$^{-1}$
over 0.324~deg$^2$ with the beam size of 2.\arcsec 8$\times$1.\arcsec 6.
\citet[][]{Miller2013} provide a catalog that includes only the objects 
above 5~$\sigma$, which corresponds to 37.0~$\mu$Jy. The data from the latter 
were obtained at the Australia Telescope, which have reached the rms of
14~$\mu$Jy~beam$^{-1}$ over 3.6~deg$^2$ with the beam size of
16\arcsec$\times$7\arcsec. The catalog also only includes the objects above
5~$\sigma$ ($\gtrsim 70\mu$Jy). Their positional uncertainties contributing to
the total positional errors are therefore
$\mathrm{\sigma_{pos}^{VLA}}=$~0.\arcsec 39 and 
$\mathrm{\sigma_{pos}^{AT}}=$~2.\arcsec 10, respectively.

    Two of our objects are within the coverage of \citet[][]{Miller2013}, but
neither has a counterpart in their catalog. We assign them the 5~$\sigma$ upper
limit of 37~$\mu$Jy. 

    There are 32 additional objects that fall within the coverage of 
\citet[][]{Franzen2015}, and 8 of them are matched with counterparts in
their catalog. In particular, \texttt{500R\_CDFS\_T1\_x004} has
$S_{1.4}=10.86\pm 0.56$~mJy, which indicates that its radio emission is most
likely powered by AGN. The next brightest, \texttt{x183}, has 
$S_{1.4}=0.95\pm 0.06$~mJy, is a borderline case. The other six objects
are likely star formation dominated. For the 24 objects that are not matched,
we assign them the 5~$\sigma$ upper limit of 70~$\mu$Jy.

   --- {\it COSMOS}\,\,\, This field has the best radio data among all,
especially those obtained at the VLA. The 1.4~GHz source catalogs are presented
in \citet[][]{Schinnerer2007, Schinnerer2010}. A deeper radio map at 3~GHz has
been obtained by \citet[][]{Smolcic2017}, which covers 2~deg$^2$ to rms of
2.3~$\mu$Jy~beam$^{-1}$ at the resolution of $0.\arcsec75$ (corresponding to
$\mathrm{\sigma_{pos}^{VLA3GHz}}=0$.\arcsec 13). The source catalog includes
10,830 sources above 5~$\sigma$ ($\gtrsim$ 11.5~$\mu$Jy).
Here we only use the 3~GHz data because they are deeper and also cover a wider
area than the 1.4~GHz data. 

    There are 17 of our objects falling within the 3~GHz coverage, among which
10 have counterparts with $\Delta/\sigma\leq 1.5$. All of these counterparts
have $S_{3.0}<100$~$\mu$Jy, with the faintest having 13~$\mu$Jy. It is highly
likely that their radio fluxes are powered by star formation. The other seven
objects are not matched (three of them have radio sources within 10\arcsec\
but have $\Delta/\sigma>1.5$), for which we assign the 5~$\sigma$ $S_{3.0}$ 
upper limit of 11.5~$\mu$Jy. Assuming a power-law SED of 
$S_\nu\sim\nu^{-0.8}$, this corresponds to $S_{1.4}$ upper limit of 
21.2~$\mu$Jy.

  --- {\it EGS}\,\,\, The deepest public radio data in this field are those
from the VLA 1.4~GHz survey of \citet[][]{Ivison2007a}, which have reached rms
of $\sim 26$~$\mu$Jy~beam$^{-1}$ over 0.73~deg$^2$ with the beam size of
$\sim$ 3.\arcsec 8 (corresponding to 
$\mathrm{\sigma_{pos}^{VLA}}=0$.\arcsec 66).
Eight of our objects are within the coverage, and only two
of them are matched with counterparts in their catalog. They both have
$S_{1.4}<100$~$\mu$Jy and are likely star formation dominated.
For the six unmatched, we assign a universal 5~$\sigma$ upper limit of 
130~$\mu$Jy, which are all in the star-formation-dominated regime. 

  --- {\it ELAISN1}\,\,\, The only radio resources appropriate here are the
data taken by the Giant Meterwave Radio Telescope (GMRT) at 610~MHz
\citet[][]{Garn2008a}, which cover $\sim 9$~deg$^2$ to an rms of
$\sim 40$-70~$\mu$Jy~beam$^{-1}$ with the beam size of
6\arcsec$\times$5\arcsec\ (corresponding to 
$\mathrm{\sigma_{pos}^{GMRT}}=0$.\arcsec 94).
The source catalog includes the objects with 610~MHz flux density 
$S_{610}>0.14$~mJy. 

    Thirty-five of our objects are covered by this radio map, but none of them
are matched with counterparts in the aforementioned 610~MHz catalog. We opt
to assign a rather conservative upper limit of 0.35~mJy (about 5~$\sigma$
in the less sensitive region of the map) to these sources. Assuming a power-law
SED of $S_{\nu}\sim \nu^{-0.8}$, this corresponds to $S_{1.4}<0.18$~mJy, which
is still in the sub-millijansky regime and thus the sources are likely star 
formation dominated.

  --- {\it ELAISN2}\,\,\, The only appropriate radio data in this field are
also obtained by the GMRT at 610~MHz \citet[][]{Garn2009}, which cover 
$\sim 6$~deg$^2$ to rms $\sim 85$~$\mu$Jy~beam$^{-1}$ with the beam size of 
6.\arcsec 5$\times$5\arcsec\ (corresponding to
$\mathrm{\sigma_{pos}^{GMRT}}=0$.\arcsec 98). The source catalog includes
the objects with $S_{610}>0.3$~mJy.

    Thirty-one of our objects are within the coverage of this radio map. Two
of them are matched with counterparts, and both are FIRST sources as already
discussed above. For the other 29 objects, we assign a conservative upper limit
of 0.425~mJy (corresponding to $S_{1.4}<0.22$~mJy when assuming
$S_{\nu}\sim \nu^{-0.8}$). Similar to the discussion above, these objects are
likely star formation dominated in radio.
    
  --- {\it ELAISS1}\,\,\, This field has 1.4~GHz data from the Australia
Telescope as described in \citet[][]{Franzen2015}, which cover 2.7~deg$^2$ to
17~$\mu$Jy~beam$^{-1}$ with the beam size of 16\arcsec$\times$7\arcsec
(corresponding to $\mathrm{\sigma_{pos}^{AT}}=2$.\arcsec 10). The public source 
catalog includes the sources that are above 5~$\sigma$. Sixteen of our objects
are within the coverage; however, none of them are matched with counterparts.
We therefore assign them the 5~$\sigma$ upper limit of 75~$\mu$Jy.

  --- {\it FLS}\,\,\, About 5~deg$^2$ of this field have 1.4~GHz data obtained 
by the VLA to rms $\sim 23$~$\mu$Jy~beam$^{-1}$ with the beam size of 5\arcsec\,
(corresponding to $\mathrm{\sigma_{pos}^{VLA}}=0$.\arcsec 86). These are 
described in \citet[][]{Condon2003}, who also present a catalog that includes 
the sources above 5~$\sigma$. Their 1.4~GHz map covers 27 of our objects, among
which 6 are matched in the radio catalog within 10\arcsec. Three of them have
already been discussed in the match to the FIRST sources: 
\texttt{500R\_FLS\_T1\_x07} and \texttt{x50} have reliable FIRST counterparts,
while \texttt{R11} does not. The match to \texttt{R11} in this catalog also has
$\Delta/\sigma>1.5$ (it is the same radio source as in the FIRST catalog) and
therefore is not a reliable counterpart. The other three new matches here are
all reliable ones and are likely dominated by star formation. For the 21
unmatched objects, we assign the 5~$\sigma$ upper limit of 115~$\mu$Jy.

  --- {\it GOODS-N}\,\,\, This field has deep VLA 1.4~GHz data as described in
\citet[][]{Morrison2010}, who cover its central 40\arcmin$\times$40\arcmin\,
area, with the beam size of $\sim$ 1.\arcsec 7 (corresponding to
$\mathrm{\sigma_{pos}^{VLA}}=0$.\arcsec 29). The data have reached an rms
noise of $\sim 3.9$~$\mu$Jy~beam$^{-1}$ near the center and 
$\sim 8$~$\mu$Jy~beam$^{-1}$ at 15\arcmin\, from the center. The catalog of
\citet[][]{Morrison2010} include sources stronger than 20~$\mu$Jy.

    Four of our objects are in the radio coverage. Two are matched within 
10\arcsec, among which only one is reliable and is likely dominated by star
formation. For those that are not matched (including the one without a reliable
match), we assign an upper limit of 20~$\mu$Jy.

  -- {\it LH}\,\,\, While this field has a number of deep radio surveys 
\citep[see, e.g., ][]{Ibar2009, Owen2008, Owen2009, Vernstrom2016} within 
limited areas, unfortunately none of our objects fall within their coverages.
The medium-deep GMRT 610~MHz surveys of \citet[][]{Garn2008b, Garn2010} offer
the only radio maps that cover most of the field (and thus most of our objects)
to a sensitivity level that is significantly deeper than FIRST. The combined
610~MHz map extends $\sim 13$~deg$^2$ to rms $\sim 80$~$\mu$Jy~beam$^{-1}$ with
the beam size of 6\arcsec$\times$5\arcsec\ (corresponding to 
$\mathrm{\sigma_{pos}^{GMRT}}=0$.\arcsec 94). The central $\sim 5$~deg$^2$
region has reached a slightly deeper sensitivity of rms 
$\sim 60$~$\mu$Jy~beam$^{-1}$ \citep[][]{Garn2008b}. The catalogs presented in 
these two papers, which are extracted independently, include the sources above
$S_{610}>0.2$~mJy. For the matching here, we merge the two catalogs into one
that contains unique sources.

   In total, 89 of our objects falling within this 610~MHz map. Only three of
them are matched within 10\arcsec\ and are all reliable matches. One of them
(\texttt{500R\_LH\_T1\_x003}) is a FIRST source already mentioned above. The
other two have $S_{610}=1.13\pm 0.12$ and $0.546\pm 0.08$~mJy, respectively.
If assuming a power-law SED of $S(\nu)\sim \nu^{-0.8}$ they would have
$S_{1.4}=0.58$ and 0.28~mJy, respectively, which would be in the AGN-dominated 
regime. We assign a conservative $S_{610}$ upper limit of 0.3~mJy to the 86
unmatched objects, which would be in the sub-millijansky regime at 1.4~GHz 
($S_{1.4}<0.15$~mJy) and hence are likely star formation dominated.
    
  -- {\it XMM-LSS}\,\,\, This field has VLA 1.4~GHz data covering
$\sim 1.3$~deg$^2$ as described in \citet[][]{Simpson2006}. However, their
source catalog only covers 0.8~deg$^2$ and includes the objects that have
$S_{1.4}>100$~$\mu$Jy. The beam size is 5\arcsec$\times$4\arcsec\,
(corresponding to $\mathrm{\sigma_{pos}^{VLA}}=0$.\arcsec 77). Only one of our
objects is within the coverage, and it is not matched with a counterpart. We
assign an upper limit of 100~$\mu$Jy, which is in the star formation dominated
regime.

    The above results are detailed in Tables \ref{tab:500RtoDeepRadio} and
\ref{tab:500RnonRadio} for the radio detections and nondetections, respectively.

\startlongtable
\begin{deluxetable*}{rccc|rccc|rccc}
\tablecolumns{12}
\tabletypesize{\scriptsize}
\tablecaption{Tier 1 500~$\mu$m risers undetected in radio
    \label{tab:500RnonRadio}}
\tablehead{
\colhead{ID} &
\colhead{$f_{500}$} & 
\colhead{$S\mathrm{_{radio}}$} & 
\colhead{Hi$z$Idx} &
\colhead{ID} &
\colhead{$f_{500}$} & 
\colhead{$S\mathrm{_{radio}}$} & 
\colhead{Hi$z$Idx} &
\colhead{ID} &
\colhead{$f_{500}$} & 
\colhead{$S\mathrm{_{radio}}$} & 
\colhead{Hi$z$Idx} \\
\colhead{} &
\colhead{(mJy)} &
\colhead{($\mu$Jy)} &
\colhead{(500)} &
\colhead{} &
\colhead{(mJy)} &
\colhead{($\mu$Jy)} &
\colhead{(500)} &
\colhead{} &
\colhead{(mJy)} &
\colhead{($\mu$Jy)} &
\colhead{(500)}
}
\startdata
    500R\_ADFS\_T1\_x03 & $56.5\pm6.0$ & $<1000$ & $>0.06$  & x28 & $34.0\pm5.8$ & $<1000$ & $>0.03$  & R20 &  $33.8\pm5.5$ & $<1000$ & $>0.03$ \\
                    x14 & $29.6\pm5.6$ & $<1000$ & $>0.03$  & x41 &  $37.1\pm5.4$ & $<1000$ & $>0.04$ &  R34 &  $28.3\pm5.5$ & $<1000$ & $>0.03$ \\
                    x18 & $37.9\pm5.5$ & $<1000$ & $>0.04$  & R05 &  $42.0\pm5.5$ & $<1000$ & $>0.04$ &      &               &         &         \\
                    x20 & $37.1\pm5.7$ & $<1000$ & $>0.04$  & R14 &  $35.3\pm5.5$ & $<1000$ & $>0.04$ &      &               &         &         \\
\hline
   500R\_CDFS\_T1\_x007 & $45.4\pm3.9$ & $<70$   & $>0.65$  & x124 & $33.5\pm4.1$ & $<70$ & $>0.48$ & x211 &  $31.1\pm3.9$ & $<70$ & $>0.44$ \\
                   x015 & $56.2\pm3.9$ & $<70$   & $\mathbf{>0.80^{*}}$  & x134 & $22.7\pm3.9$ & $<70$ & $>0.32$ & x240 &  $23.9\pm4.0$ & $<70$ & $>0.34$ \\
                   x034 & $44.7\pm4.3$ & $<70$   & $>0.64$  & x166 & $22.3\pm4.2$ & $<70$ & $>0.32$ & x246 &  $36.6\pm4.6$ & $<70$ & $>0.52$ \\
                   x037 & $36.2\pm5.4$ & $<70$   & $>0.52$  & x170 & $20.9\pm4.1$ & $<70$ & $>0.30$ & x268 &  $21.4\pm4.0$ & $<70$ & $>0.31$ \\
                   x038 & $37.7\pm4.0$ & $<70$   & $>0.54$  & x176 & $27.2\pm3.9$ & $<70$ & $>0.39$ & x284 &  $32.1\pm4.0$ & $<70$ & $>0.46$ \\
                   x048 & $27.9\pm4.2$ & $<37$   & $\mathbf{>0.75^{*}}$  & x178 & $30.2\pm4.0$ & $<70$ & $>0.43$ & R14  &  $26.8\pm3.9$ & $<70$ & $>0.38$ \\
                   x058 & $25.9\pm4.2$ & $<70$   & $>0.37$  & x180 & $28.9\pm4.8$ & $<70$ & $>0.41$ & R27  &  $22.8\pm4.0$ & $<70$ & $>0.33$ \\
                   x118 & $27.9\pm3.9$ & $<70$   & $>0.40$  & x189 & $29.1\pm4.0$ & $<70$ & $>0.42$ & R32  &  $22.0\pm4.0$ & $<70$ & $>0.31$ \\
                   x119 & $25.8\pm4.2$ & $<70$   & $>0.37$  & x190 & $28.6\pm4.0$ & $<37$ & $\mathbf{>0.77^{*}}$ &      &               &       &         \\
\hline
  500R\_COSMOS\_T1\_x02 & $26.4\pm5.0$ & $<11.5$ & $\mathbf{>1.25^{*}}$  & x16 & $28.9\pm5.0$ & $<11.5$ & $\mathbf{>1.37^{*}}$ & x35 & $41.0\pm5.0$ & $<11.5$ & $\mathbf{>1.94^{*}}$ \\
                    x04 & $27.2\pm5.1$ & $<11.5$ & $\mathbf{>1.29^{*}}$  & x21 & $30.8\pm5.0$ & $<11.5$ & $\mathbf{>1.45^{*}}$ &      &               &       &         \\
                    x10 & $28.3\pm5.2$ & $<11.5$ & $\mathbf{>1.34^{*}}$  & x34 & $39.8\pm5.1$ & $<11.5$ & $\mathbf{>1.87^{*}}$ &      &               &       &         \\
\hline
     500R\_EGS\_T1\_x08 & $33.0\pm4.4$ & $<130$ & $>0.25$  & x36 & $29.1\pm4.3$ & $<130$ & $>0.22$ & x45 & $23.3\pm4.3$ & $<130$ & $>0.18$ \\
                    x09 & $25.0\pm4.6$ & $<130$ & $>0.19$  & x42 & $27.1\pm4.5$ & $<130$ & $>0.21$ & x46 & $24.0\pm4.8$ & $<130$ & $>0.18$ \\
\hline
  500R\_ELAISN1\_T1\_x001 & $170.7\pm5.1$ & $<350$ & $\mathbf{>0.95^{*}}$ & x044 & $35.0\pm5.0$ & $<350$ & $>0.19$ & x075 & $38.9\pm5.1$ & $<350$ & $>0.22$ \\
                     x003 & $51.1\pm5.3$ & $<350$ & $>0.28$ & x045 & $39.5\pm5.2$ & $<350$ & $>0.22$ & x078 & $40.0\pm5.8$ & $<350$ & $>0.22$ \\
                     x004 & $55.2\pm4.9$ & $<350$ & $>0.31$ & x053 & $28.4\pm5.0$ & $<350$ & $>0.16$ & x079 & $36.0\pm5.1$ & $<350$ & $>0.20$ \\
                     x008 & $55.2\pm4.9$ & $<350$ & $>0.31$ & x054 & $48.9\pm5.0$ & $<350$ & $>0.27$ & x081 & $29.8\pm5.2$ & $<350$ & $>0.17$ \\
                     x009 & $43.1\pm5.0$ & $<350$ & $>0.24$ & x056 & $28.9\pm5.1$ & $<350$ & $>0.16$ & x082 & $28.2\pm5.2$ & $<350$ & $>0.16$ \\
                     x012 & $31.1\pm5.1$ & $<350$ & $>0.17$ & x057 & $27.6\pm5.2$ & $<350$ & $>0.15$ & x083 & $29.3\pm5.0$ & $<350$ & $>0.16$ \\
                     x025 & $36.9\pm5.7$ & $<350$ & $>0.21$ & x060 & $33.8\pm5.2$ & $<350$ & $>0.19$ & x087 & $27.9\pm5.5$ & $<350$ & $>0.16$ \\
                     x026 & $37.6\pm5.0$ & $<350$ & $>0.21$ & x061 & $40.8\pm5.2$ & $<350$ & $>0.23$ & x091 & $28.6\pm5.3$ & $<350$ & $>0.16$ \\
                     x028 & $27.4\pm5.0$ & $<350$ & $>0.15$ & x062 & $32.9\pm5.1$ & $<350$ & $>0.18$ & x093 & $34.0\pm4.9$ & $<350$ & $>0.19$ \\
                     x031 & $32.6\pm4.9$ & $<350$ & $>0.18$ & x064 & $35.7\pm5.9$ & $<350$ & $>0.20$ & x099 & $25.3\pm4.9$ & $<350$ & $>0.14$ \\
                     x033 & $36.5\pm5.5$ & $<350$ & $>0.20$ & x068 & $28.0\pm5.0$ & $<350$ & $>0.16$ & x103 & $25.9\pm5.0$ & $<350$ & $>0.14$ \\
                     x038 & $35.8\pm4.9$ & $<350$ & $>0.20$ & x070 & $40.9\pm5.2$ & $<350$ & $>0.23$ &      &               &       &         \\
\hline
   500R\_ELAISN2\_T1\_x02 & $65.8\pm6.1$ & $<425$ & $>0.30$ & x27 &  $45.0\pm5.9$ & $<425$ & $>0.21$ & R05 &  $46.9\pm5.8$ & $<425$ & $>0.21$ \\
                      x04 & $52.5\pm5.9$ & $<425$ & $>0.24$ & x28 &  $33.6\pm6.4$ & $<425$ & $>0.15$ & R06 &  $44.7\pm6.0$ & $<425$ & $>0.20$ \\
                      x07 & $48.8\pm5.9$ & $<425$ & $>0.22$ & x30 &  $33.4\pm6.0$ & $<425$ & $>0.15$ & R07 &  $43.3\pm5.9$ & $<425$ & $>0.20$ \\
                      x08 & $48.3\pm6.2$ & $<425$ & $>0.22$ & x31 &  $35.0\pm6.2$ & $<425$ & $>0.16$ & R08 &  $43.0\pm5.9$ & $<425$ & $>0.20$ \\
                      x09 & $42.7\pm6.2$ & $<425$ & $>0.20$ & x33 &  $35.8\pm6.1$ & $<425$ & $>0.16$ & R15 &  $37.3\pm5.8$ & $<425$ & $>0.17$ \\
                      x12 & $44.4\pm6.0$ & $<425$ & $>0.20$ & x36 &  $53.8\pm7.1$ & $<425$ & $>0.25$ & R17 &  $35.0\pm5.9$ & $<425$ & $>0.16$ \\
                      x18 & $31.2\pm6.0$ & $<425$ & $>0.14$ & x38 &  $34.9\pm5.9$ & $<425$ & $>0.16$ & R27 &  $31.4\pm5.9$ & $<425$ & $>0.14$ \\
                      x19 & $38.2\pm6.0$ & $<425$ & $>0.17$ & x43 &  $32.8\pm6.0$ & $<425$ & $>0.15$ & R28 &  $29.8\pm5.8$ & $<425$ & $>0.14$ \\
                      x20 & $38.2\pm5.9$ & $<425$ & $>0.17$ & x45 &  $31.4\pm6.0$ & $<425$ & $>0.14$ & R29 &  $29.6\pm5.9$ & $<425$ & $>0.14$ \\
                      x24 & $32.6\pm5.9$ & $<425$ & $>0.15$ & x51 &  $34.1\pm6.4$ & $<425$ & $>0.16$ &      &               &       &         \\
\hline
   500R\_ELAISS1\_T1\_x02 & $44.8\pm4.8$ & $<75$ & $>0.60$  & x47 &  $32.0\pm4.7$ & $<75$ & $>0.43$  & x84 & $25.9\pm4.6$ & $<75$ & $>0.35$ \\
                      x18 & $53.3\pm4.7$ & $<75$ & $\mathbf{>0.71^{*}}$  & x55 &  $26.6\pm4.7$ & $<75$ & $>0.35$  & x85 & $25.4\pm4.7$ & $<75$ & $>0.34$ \\
                      x26 & $37.2\pm5.0$ & $<75$ & $>0.50$  & x61 &  $24.0\pm4.7$ & $<75$ & $>0.32$  & R16 & $29.0\pm4.6$ & $<75$ & $>0.39$ \\
                      x37 & $26.6\pm4.9$ & $<75$ & $>0.35$  & x71 &  $25.2\pm4.7$ & $<75$ & $>0.34$  & R26 & $25.8\pm5.1$ & $<75$ & $>0.34$ \\
                      x45 & $28.2\pm4.7$ & $<75$ & $>0.38$  & x72 &  $35.3\pm4.8$ & $<75$ & $>0.47$  &      &               &       &         \\
                      x46 & $25.0\pm4.8$ & $<75$ & $>0.33$  & x76 &  $28.1\pm5.0$ & $<75$ & $>0.37$  &      &               &       &         \\
\hline
        500R\_FLS\_T1\_x02 & $41.0\pm5.9$ & $<150$ & $>0.27$ & x23 &  $33.2\pm6.1$ & $<150$ & $>0.22$ & R01 & $51.0\pm5.8$ & $<150$ & $>0.34$ \\ 
                      x09 & $48.7\pm5.8$ & $<150$ & $>0.32$ & x29 &  $32.8\pm6.0$ & $<150$ & $>0.22$ & R02 & $46.5\pm5.8$ & $<150$ & $>0.31$ \\ 
                      x12 & $33.7\pm5.8$ & $<150$ & $>0.22$ & x31 &  $64.4\pm5.8$ & $<150$ & $>0.43$ & R05 & $39.9\pm5.8$ & $<150$ & $>0.27$ \\ 
                      x15 & $41.9\pm5.8$ & $<150$ & $>0.28$ & x32 &  $32.5\pm5.8$ & $<150$ & $>0.22$ & R07 & $38.6\pm5.8$ & $<150$ & $>0.26$ \\ 
                      x16 & $38.2\pm5.9$ & $<150$ & $>0.25$ & x33 &  $30.2\pm5.8$ & $<150$ & $>0.20$ & R10 & $36.1\pm5.8$ & $<150$ & $>0.24$ \\ 
                      x18 & $45.0\pm5.9$ & $<150$ & $>0.30$ & x37 &  $34.8\pm5.8$ & $<150$ & $>0.23$ & R13 & $32.5\pm5.8$ & $<150$ & $>0.22$ \\ 
                      x21 & $47.2\pm5.9$ & $<150$ & $>0.31$ & x46 &  $33.8\pm6.6$ & $<150$ & $>0.23$ & R16 & $29.1\pm5.8$ & $<150$ & $>0.19$ \\ 
\hline
       500R\_GOODSN\_T1\_x01 & $43.5\pm3.5$ & $<20.0$ & $\mathbf{>2.18^{*}}$ & x03 & $29.2\pm3.4$ & $<20.0$ & $\mathbf{>1.46^{*}}$ & x09 & $24.1\pm3.4$ & $<20.0$ & $\mathbf{>1.20^{*}}$ \\
\hline
   500R\_LH\_T1\_x001 & $137.5\pm4.5$ & $<300$ & $>0.46$ &  x126 & $30.9\pm4.5$ & $<300$ & $>0.10$ &  x230 & $30.3\pm4.6$ & $<300$ & $>0.10$ \\
                     x002 & $75.2\pm0.0$ & $<300$ & $>0.25$ &  x127 & $32.8\pm4.6$ & $<300$ & $>0.11$ &  x233 & $35.6\pm4.6$ & $<300$ & $>0.12$ \\ 
                     x005 & $70.8\pm4.8$ & $<300$ & $>0.24$ &  x136 & $24.7\pm4.6$ & $<300$ & $>0.08$ &  x235 & $26.1\pm5.2$ & $<300$ & $>0.09$ \\ 
                     x012 & $50.2\pm4.6$ & $<300$ & $>0.17$ &  x137 & $29.1\pm4.7$ & $<300$ & $>0.10$ &  x238 & $25.3\pm4.5$ & $<300$ & $>0.08$ \\ 
                     x019 & $43.4\pm5.3$ & $<300$ & $>0.14$ &  x138 & $38.3\pm4.5$ & $<300$ & $>0.13$ &  x239 & $23.2\pm4.5$ & $<300$ & $>0.08$ \\ 
                     x038 & $42.0\pm4.7$ & $<300$ & $>0.14$ &  x154 & $25.0\pm4.6$ & $<300$ & $>0.08$ &  x240 & $24.8\pm4.8$ & $<300$ & $>0.08$ \\ 
                     x041 & $31.4\pm4.6$ & $<300$ & $>0.10$ &  x155 & $22.6\pm4.5$ & $<300$ & $>0.08$ &  x245 & $30.5\pm4.6$ & $<300$ & $>0.10$ \\ 
                     x056 & $45.1\pm4.5$ & $<300$ & $>0.15$ &  x162 & $29.9\pm4.7$ & $<300$ & $>0.10$ &  x248 & $31.5\pm4.7$ & $<300$ & $>0.11$ \\ 
                     x057 & $36.2\pm5.6$ & $<300$ & $>0.12$ &  x170 & $27.3\pm4.5$ & $<300$ & $>0.09$ &  x249 & $30.7\pm4.9$ & $<300$ & $>0.10$ \\ 
                     x058 & $34.8\pm4.5$ & $<300$ & $>0.12$ &  x172 & $29.1\pm4.6$ & $<300$ & $>0.10$ &  x252 & $26.1\pm4.7$ & $<300$ & $>0.09$ \\ 
                     x062 & $29.9\pm4.6$ & $<300$ & $>0.10$ &  x173 & $32.4\pm5.2$ & $<300$ & $>0.11$ &   R11 & $46.6\pm4.6$ & $<300$ & $>0.16$ \\ 
                     x063 & $27.5\pm4.7$ & $<300$ & $>0.09$ &  x174 & $25.6\pm4.9$ & $<300$ & $>0.09$ &   R13 & $46.3\pm5.9$ & $<300$ & $>0.15$ \\ 
                     x067 & $34.8\pm4.8$ & $<300$ & $>0.12$ &  x179 & $25.5\pm4.8$ & $<300$ & $>0.09$ &   R22 & $37.3\pm4.7$ & $<300$ & $>0.12$ \\ 
                     x070 & $40.5\pm4.7$ & $<300$ & $>0.14$ &  x180 & $26.7\pm4.5$ & $<300$ & $>0.09$ &   R27 & $36.1\pm5.2$ & $<300$ & $>0.12$ \\ 
                     x076 & $42.9\pm4.5$ & $<300$ & $>0.14$ &  x184 & $27.7\pm4.6$ & $<300$ & $>0.09$ &   R37 & $33.8\pm5.5$ & $<300$ & $>0.11$ \\ 
                     x078 & $29.3\pm5.2$ & $<300$ & $>0.10$ &  x188 & $26.0\pm4.6$ & $<300$ & $>0.09$ &   R39 & $33.1\pm5.2$ & $<300$ & $>0.11$ \\ 
                     x080 & $30.1\pm4.7$ & $<300$ & $>0.10$ &  x190 & $25.7\pm4.5$ & $<300$ & $>0.09$ &   R43 & $31.2\pm4.5$ & $<300$ & $>0.10$ \\ 
                     x085 & $36.0\pm4.7$ & $<300$ & $>0.12$ &  x192 & $25.4\pm4.6$ & $<300$ & $>0.08$ &   R45 & $31.0\pm4.8$ & $<300$ & $>0.10$ \\ 
                     x090 & $30.0\pm4.5$ & $<300$ & $>0.10$ &  x196 & $29.8\pm4.5$ & $<300$ & $>0.10$ &   R50 & $30.2\pm4.8$ & $<300$ & $>0.10$ \\ 
                     x092 & $28.6\pm4.6$ & $<300$ & $>0.10$ &  x199 & $23.2\pm4.5$ & $<300$ & $>0.08$ &   R51 & $29.7\pm4.7$ & $<300$ & $>0.10$ \\ 
                     x093 & $39.4\pm5.1$ & $<300$ & $>0.13$ &  x201 & $23.0\pm4.5$ & $<300$ & $>0.08$ &   R55 & $27.3\pm4.7$ & $<300$ & $>0.09$ \\ 
                     x096 & $30.7\pm4.6$ & $<300$ & $>0.10$ &  x203 & $25.6\pm4.5$ & $<300$ & $>0.09$ &   R57 & $26.6\pm4.6$ & $<300$ & $>0.09$ \\ 
                     x097 & $30.9\pm4.6$ & $<300$ & $>0.10$ &  x206 & $27.9\pm5.2$ & $<300$ & $>0.09$ &   R58 & $26.5\pm4.6$ & $<300$ & $>0.09$ \\ 
                     x099 & $29.7\pm4.5$ & $<300$ & $>0.10$ &  x208 & $31.5\pm4.7$ & $<300$ & $>0.10$ &   R66 & $25.1\pm4.8$ & $<300$ & $>0.08$ \\ 
                     x105 & $35.0\pm4.8$ & $<300$ & $>0.12$ &  x211 & $26.0\pm4.5$ & $<300$ & $>0.09$ &   R67 & $25.1\pm4.6$ & $<300$ & $>0.08$ \\ 
                     x107 & $35.3\pm4.8$ & $<300$ & $>0.12$ &  x216 & $24.9\pm4.5$ & $<300$ & $>0.08$ &   R68 & $24.9\pm4.6$ & $<300$ & $>0.08$ \\ 
                     x110 & $35.2\pm4.7$ & $<300$ & $>0.12$ &  x217 & $29.4\pm4.8$ & $<300$ & $>0.10$ &   R71 & $24.3\pm4.8$ & $<300$ & $>0.08$ \\ 
                     x119 & $29.7\pm4.9$ & $<300$ & $>0.10$ &  x224 & $22.7\pm4.5$ & $<300$ & $>0.08$ &   R73 & $24.2\pm4.6$ & $<300$ & $>0.08$ \\ 
                     x125 & $33.6\pm5.4$ & $<300$ & $>0.11$ &  x226 & $30.3\pm4.6$ & $<300$ & $>0.10$ &       &              &        &         \\        
\hline
    500R\_XMMLSS\_T1\_x70 & $39.4\pm6.5$ & $<100$ & $>0.39$ &       &              &        &         &       &              &        &         \\
\enddata
\tablecomments{Tier 1 500~$\mu$m risers covered by various radio surveys
deeper than FIRST but are not detected. As in Table \ref{tab:500RCatSample},
the leading string in the object ID is omitted for clarity except when the
object is the first entry of a given field. The upper limits of their radio flux 
densities are based on the radio catalogs described in \S \ref{sec:500R_Radio}.
The lower limits of Hi$z$Idx(500) are calculated using these upper limits. Those
that have Hi$z$Idx(500) $\geq 0.7$ (potentially at $z>6$) are boldfaced and marked
with asterisk (see \S \ref{sec:highz}).
}
\end{deluxetable*}

\subsubsection{Summary of radio source matching}

   In most cases, the currently available radio data in these fields are still
very limited in both coverage and sensitivity. Nevertheless, we can still
draw two important conclusions. First, radio data are effective in revealing
the contamination from blazars whose SEDs have the same rising trend but are
due to their nonthermal emissions. While they only constitute a small fraction
($\sim 1.0$\%) among 500~$\mu$m risers (see Table 5), these contaminators are 
often the brightest among the sample and thus are the easiest to select
for follow-up studies. Therefore, checking the sample against radio surveys 
sensitive to a few mJy level will be desirable. Second, only a small fraction 
of 500~$\mu$m risers harbor nonblazar, radio-loud AGNs. The FIRST data, whose
sensitivity reaches well below $\sim 1$~mJy (the conventional dividing line
between AGN-dominated and star-formation-dominated regimes), have revealed 
three such AGNs, or $\sim 0.8$\%, among our 500~$\mu$m risers (see Table 5).
The data in the CDFS (not accessible by FIRST) have added two more (including 
one at the borderline; see Table 6) among the 32 that are within the radio
coverage. If adding the latter, the radio-loud AGN fraction among our 
500~$\mu$m risers is $\sim 1.2$\%. 

  We shall point out that harboring a radio-loud AGN is not a criterion
against a source being at high redshift. In fact, AGNs as strong as quasars at
high redshifts can have cold-dust emissions due to heating by star formation,
and it is known that such systems do exist \citep[see, e.g.,][]{MY15}. Due to 
the lack of further observations, we will not discuss these radio-loud AGNs 
any further in this paper. On the other hand, 30 of our 500~$\mu$m risers have
radio emissions in the star-formation-dominated regime (\S 6.1.2). As we will
discuss in \S \ref{sec:highz}, they are not likely at high redshifts. The more
interesting objects are those that are weak or not detected in the deepest
radio data, which will be discussed in detail in \S \ref{sec:highz}.

\subsection{Radio Properties of SPIRE Dropouts\label{sec:SD850_Radio}}

   For the same reason, here we discuss the radio properties of the 
SPIRE dropouts in the areas where sufficiently deep radio data are available.
We again use the same matching procedure as in \S \ref{sec:500R_Radio}, i.e.,
searching for matches within the matching radius of $r=10$\arcsec, and then
determining if a match is reliable by checking if it satisfies 
$\Delta/\sigma\leq 1.5$. The dominant term to $\mathrm{\sigma_{pos}}$ is 
$\mathrm{\sigma_{SCUBA2}}$, which is the positional uncertainty due to the 
large beam size of SCUBA2 850~$\mu$m (13\arcsec) and is calculated individually
based on the source S/N, ``\texttt{detection-SNR}'' as in the S2CLS catalog.
The radio source position uncertainty is calculated at the fixed 
$\mathrm{S/N=5}$ as in \S \ref{sec:500R_Radio}, and only contributes a 
negligible amount to $\mathrm{\sigma_{pos}}$ in all cases here. The matching
results are summarized in Table \ref{tab:SD850RadioSummary}. The matched
sources are shown in Table \ref{tab:SD850toDeepRadio}, and the 
unmatched ones are indicated as such in Table \ref{tab:SD850nonRadio}. We
provide some details below for all these fields.

\begin{table*}
\centering
\caption{Summary of matching Tier 1 SPIRE dropouts to deep radio data
    \label{tab:SD850RadioSummary}}
\begin{tabular}{llcccccc}
\toprule
Field & Facility & Area    & Beam Size & $\mathrm{\sigma_{pos}^{Radio}}$ & Sensitivity & Obj Coverage & Obj Matched \\
      &          & \multicolumn{1}{c}{(deg$^2$)} & \multicolumn{1}{c}{(\arcsec)} & \multicolumn{1}{c}{(\arcsec)} & \multicolumn{1}{c}{($\mu$Jy)} & \\
\midrule
 COSMOS    & VLA, 3.0~GHz  & 1.34  & 0.75             & 0.13  & 11.5    & 26/26  & 10 \\
 EGS       & VLA, 1.4~GHz  & 0.16  & 3.8              & 0.66  & 130     &  4/6   &  0 \\
 GOODS-N   & VLA, 1.4~GHz  & 0.07  & 1.7              & 0.29  &  20     &  3/3   &  1 \\
 LH        & VLA, 1.4~MHz  & 0.28  & 1.63$\times$1.57 & 0.27  & 13.5    &  3/3   &  2 \\
 UDS       & VLA, 1.4~GHz  & 0.94  & 1.82$\times$1.63 & 0.29  &  60     & 57/57  &  9 \\
\bottomrule
\end{tabular}
\tablecomments{The quoted numbers under ``Area'' are for the overlapping 
regions of the 850~$\mu$m and the radio observations. The values of 
$\mathrm{\sigma_{pos}^{Radio}}$ are calculated based on Equation 
\ref{eq:poserr} and $\mathrm{S/N=5}$. The sensitivity levels are based on the
source catalogs that we used, and in most cases correspond to $\mathrm{S/N=5}$,
except in UDS where it corresponds to $\mathrm{S/N=3}$. For the 3~GHz data in
COSMOS, the conversion to 1.4~GHz for the quoted sensitivity levels can done by
assuming a power law SED of $S_{\nu} \sim \nu^{-0.8}$. The number pairs under
``Obj Coverage'' are those of the objects in the areas covered by the relevant
radio data and the total in the Tier 1 samples, respectively, while the numbers
under `` Obj Matched'' are those of the objects detected above the quoted
sensitivity levels.
}
\end{table*}

   -- {\it COSMOS}\,\,\, The VLA 3~GHz map of \citet[][]{Smolcic2017} covers
the entire field and hence all of our SPIRE dropouts in this field. Among the
26 Tier 1 objects, 15 objects have 3~GHz matches within $r=10$\arcsec. Ten of
them have $\Delta/\sigma\leq 1.5$ and hence are reliable. The other five
are not regarded as the real counterparts. These 5 objects, together with 
the 11 that have no entries in the 3~GHz catalog, are assigned the 
conservative, 5~$\sigma$ upper limit of $S_{3.0}\leq 11.5$~$\mu$Jy
(corresponding to $S_{1.4}\leq 21.2$~$\mu$Jy under $S_{\nu}\sim \nu^{-0.8}$).

   -- {\it EGS}\,\,\, The VLA 1.4~GHz map of \citet[][]{Ivison2007a} covers
about only half of the field and four of our objects (out of six).
However, none are matched with radio counterparts. We assign them the 5~$\sigma$
upper limit of 130~$\mu$Jy. 

  -- {\it GOODS-N}\,\,\, The VLA 1.4~GHz map of \citet[][]{Morrison2010} covers
the entire field and hence all three of our Tier 1 objects. However, none of 
them are matched. We assign the 5~$\sigma$ upper limit of 20~$\mu$Jy to all of
them.

  -- {\it LH}\,\,\, This field is fully covered by the deep VLA 1.4~GHz map of
\cite[][]{Owen2008}, which extends over an area of 40\arcmin$ \times $40\arcmin\
and reaches rms $\sim 2.7$~$\mu$Jy~beam$^{-1}$ with the beam size of 
1.\arcsec 63$\times 1$.\arcsec 57 (corresponding to 
$\mathrm{\sigma_{pos}^{VLA}}=0$.\arcsec 27). The radio catalog includes the
sources that are detected above 5~$\sigma$. Two of the three Tier 1 objects are
matched with their radio counterparts in this catalog, with
$S_{1.4} = 28.5\pm 4.2$ and $47.7\pm 7.1$~$\mu$Jy. For the object that is not
included the radio catalog,  we assign the 5~$\sigma$ upper limit of 
13.5~$\mu$Jy.

  -- {\it UDS}\,\,\, For this field, we mostly use the VLA 1.4~GHz data of 
\citet[][]{Arumugam2013}, which are $2\times$ deeper than those of
\citet[][]{Simpson2006} and are of $3\times$ higher resolution (beam size
1.\arcsec 82$ \times $1.\arcsec 63, corresponding to 
$\mathrm{\sigma_{pos}^{VLA}}=0$.\arcsec 29). However, these
data are not directly available. We use instead the catalog of 
\citet[][]{Michalowski2017}, where all the S2CLS sources in this field have
been matched for radio counterparts to S/N $=3$ using the data of 
\citet[][]{Arumugam2013}.
As it turns out, 8 of our 57 objects have reported 1.4~GHz
detections that fulfill our matching criterion. They have $S_{1.4}$ ranging
from 56.2 to 639~$\mu$Jy, with S/N of 3.3 to 35.2. Based on the sensitivity
map provided in Figure 2.1 of \citet[][]{Arumugam2013}, we assign the universal
3~$\sigma$ upper limit of 60~$\mu$Jy to the rest objects that
do not have reported radio counterparts in \citet[][]{Michalowski2017}.
The two objects that are also in \citet[][]{Ikarashi2017}, our
\texttt{SD850\_UDS\_T1\_A36} and \texttt{A42}, need further explanation.
According to \citet[][]{Ikarashi2017}, \texttt{A42} (their ASXDF1100.231.1) is
detected at $\mathrm{S/N}=3.3$ in the data of \citet[][]{Arumugam2013} (but
no flux density is given). However, the catalog of \citet[][]{Michalowski2017}
based on the same data does not list it as being detected to $\mathrm{S/N}=3$.
Here we follow \citet[][]{Michalowski2017}, treat it as a nondetection, and
put it in Table \ref{tab:SD850nonRadio}.
For \texttt{A36} (their ASXDF1100.053.1), \citet[][]{Ikarashi2017} have
obtained their own VLA 6~GHz measurement of $S_{6.0}=4.5\pm1.1$~$\mu$Jy. Its
radio position is not given, however. We treat this source as a detection and
put it in Table \ref{tab:SD850toDeepRadio}. Thus, the number of 
radio-undetected SPIRE dropouts in this field is 48.

\begin{table*}
\scriptsize
\centering
\caption{Match of Tier 1 SPIRE dropouts to radio data
    \label{tab:SD850toDeepRadio}}
\begin{tabular}{lccccccccl}
\toprule
Object Name & $\mathrm{\sigma_{pos}}$ & $\mathrm{\Delta_{pos}}$ &
$\Delta$/$\sigma$ & $S_{850}$ & RA$\mathrm{_{radio}}$ & DEC$\mathrm{_{radio}}$ &
 $S\mathrm{_{radio}}$ &
Hi$z$Idx & References \\
            & \multicolumn{1}{c}{(\arcsec)} & \multicolumn{1}{c}{(\arcsec)} &
            & \multicolumn{1}{c}{(mJy)}
            &
            &
            & \multicolumn{1}{c}{($\mu$Jy)}
            & \multicolumn{1}{c}{(850)}        &          \\
\midrule
    SD850\_COSMOS\_T1\_A04 & 1.4 & 1.3 & 1.1 &$11.7\pm2.1$ &  9:59:10.33 &   2:48:55.72 & $21.7\pm2.7$ & 0.29 & (1) \\
    SD850\_COSMOS\_T1\_A11 & 1.5 & 1.8 & 0.8 & $5.4\pm1.4$ &  9:59:31.16 &   2:14:33.99 & $30.5\pm2.7$ & 0.10 & (1) \\
    SD850\_COSMOS\_T1\_A15 & 2.2 & 2.0 & 1.1 & $5.8\pm1.4$ & 10:00:16.99 &   1:43:26.79 & $16.9\pm2.3$ & 0.19 & (1) \\
    SD850\_COSMOS\_T1\_A18 & 3.0 & 2.0 & 1.5 & $5.1\pm1.3$ &  9:58:34.65 &   2:18:02.71 & $14.8\pm2.4$ & 0.19 & (1) \\
    SD850\_COSMOS\_T1\_A20 & 0.4 & 2.1 & 0.2 & $4.7\pm1.2$ & 10:00:47.10 &   2:10:16.62 & $11.4\pm2.3$ & 0.22 & (1) \\
    SD850\_COSMOS\_T1\_A23 & 2.2 & 2.1 & 1.0 & $6.3\pm1.7$ &  9:59:58.79 &   2:34:58.02 & $21.6\pm2.5$ & 0.16 & (1) \\
    SD850\_COSMOS\_T1\_A30 & 2.0 & 2.2 & 0.9 & $6.7\pm1.8$ &  9:58:43.44 &   2:45:18.14 & $15.7\pm2.5$ & 0.23 & (1) \\
    SD850\_COSMOS\_T1\_B06 & 2.7 & 1.9 & 1.5 & $6.3\pm1.4$ & 10:00:21.36 &   2:00:41.26 & $18.9\pm2.5$ & 0.18 & (1) \\
    SD850\_COSMOS\_T1\_B14 & 1.1 & 2.1 & 0.5 & $4.1\pm0.9$ & 10:00:23.26 &   2:13:44.09 & $14.1\pm2.4$ & 0.16 & (1) \\
    SD850\_COSMOS\_T1\_B16 & 1.2 & 2.1 & 0.5 & $6.3\pm1.6$ & 10:00:37.27 &   2:49:11.19 & $13.3\pm2.5$ & 0.26 & (1) \\
    SD850\_GOODSN\_T1\_B01 & 2.1 & 1.5 & 1.4 & $7.2\pm1.2$ & 12:36:31.92 &  62:17:14.70 & $26.4\pm4.9$ & 0.27 & (2) \\
        SD850\_LH\_T1\_A02 & 1.4 & 1.6 & 0.9 & $6.8\pm1.4$ & 10:45:02.00 &  59:04:03.30 & $28.5\pm4.2$ & 0.24 & (3) \\
        SD850\_LH\_T1\_A03 & 1.7 & 1.7 & 1.0 & $6.3\pm1.6$ & 10:47:20.72 &  58:51:53.00 & $47.7\pm7.1$ & 0.13 & (3) \\
       SD850\_UDS\_T1\_A09 & 1.4 & 1.8 & 0.8 & $6.4\pm1.1$ &  2:19:08.88 & $-$5:13:55.30 & $639.2\pm18.2$ & 0.01 & (4) \\
       SD850\_UDS\_T1\_A41 & 1.9 & 2.3 & 0.8 & $4.8\pm1.1$ &  2:16:41.95 & $-$5:07:04.00 & $56.2\pm15.8$ & 0.09 & (4) \\
       SD850\_UDS\_T1\_A48 & 4.0 & 2.7 & 1.5 & $4.1\pm1.0$ &  2:16:46.94 & $-$4:45:08.80 & $532.6\pm29.3$ & 0.01 & (4) \\
       SD850\_UDS\_T1\_A57 & 2.8 & 2.6 & 1.1 & $4.3\pm1.1$ &  2:17:51.14 & $-$4:48:06.70 & $101.0\pm16.5$ & 0.04 & (4) \\
       SD850\_UDS\_T1\_A62 & 2.6 & 2.8 & 1.0 & $4.0\pm1.0$ &  2:18:16.70 & $-$5:15:45.10 & $58.2\pm15.8$ & 0.07 & (4) \\
       SD850\_UDS\_T1\_B04 & 2.0 & 1.3 & 1.5 & $8.5\pm1.4$ &  2:18:07.19 & $-$4:44:13.80 & $67.7\pm18.6$ & 0.13 & (4) \\
       SD850\_UDS\_T1\_B11 & 2.2 & 1.6 & 1.3 & $6.9\pm1.2$ &  2:19:24.88 & $-$5:09:20.70 & $84.7\pm21.2$ & 0.08 & (4) \\
       SD850\_UDS\_T1\_B16 & 2.4 & 2.0 & 1.2 & $5.6\pm1.1$ &  2:17:03.08 & $-$4:43:17.80 & $119.8\pm36.4$ & 0.05 & (4) \\
       SD850\_UDS\_T1\_A36 & \nodata & \nodata & \nodata & $4.8\pm1.1$ & \nodata & \nodata & $4.6\pm1.1$ & 0.33 & (5) \\
\hline
    SD850\_COSMOS\_T1\_A03 & 3.0 & 1.3 & 2.4 & $11.8\pm1.9$ & 9:59:57.29 &   2:27:30.54 & $28.8\pm2.7$ & \nodata & (1) \\
    SD850\_COSMOS\_T1\_A19 & 5.7 & 2.1 & 2.8 & $5.1\pm1.2$ & 10:00:28.82 &   2:05:23.53 & $21.7\pm2.7$ & \nodata & (1) \\
    SD850\_COSMOS\_T1\_A22 & 5.0 & 2.1 & 2.3 & $5.5\pm1.4$ & 10:00:23.38 &   2:01:21.50 & $27.8\pm2.8$ & \nodata & (1) \\
    SD850\_COSMOS\_T1\_A26 & 9.6 & 2.1 & 4.5 & $4.0\pm0.9$ & 10:00:08.15 &   2:17:11.55 & $21.5\pm2.6$ & \nodata & (1) \\
    SD850\_COSMOS\_T1\_B05 & 3.6 & 1.7 & 2.1 & $8.0\pm1.8$ & 10:00:06.49 &   2:38:37.44 & $12.6\pm2.4$ & \nodata & (1) \\
    SD850\_GOODSN\_T1\_A02 & 7.2 & 1.6 & 4.6 & $5.9\pm1.3$ & 12:36:51.72 &  62:12:21.40 & $56.0\pm4.5$ & \nodata & (2) \\
\bottomrule
\end{tabular}
\tablecomments{Tier 1 SPIRE dropouts matched to the radio sources in various surveys.
The matching radius is $r=10$\arcsec. Only the top 21 matches that have 
$\Delta/\sigma\leq 1.5$ are deemed to be reliable. ``Hi$z$Idx'' is calculated based on Equation
\ref{eq:highzindex}. For the bottom six objects, they only have lower limits of Hi$z$Idx because
they are treated as with no radio counterparts and thus only have upper limits of their radio flux
densities.
The radio data are from the references listed in the last column. These are:
(1) \citet[][]{Smolcic2017}, 3~GHz; (2) \citet[][]{Morrison2010}, 1.4~GHz; (3) \citet[][]{Owen2008}, 1.4~GHz; 
(4) \citet[][]{Michalowski2017}, 1.4~GHz; (5) \citet[][]{Ikarashi2017}, 6~GHz. 
}
\end{table*}


\begin{deluxetable*}{rccc|rccc|rccc}
\tablecolumns{12}
\tabletypesize{\scriptsize}
\tablecaption{Tier 1 SPIRE dropouts undetected in radio
    \label{tab:SD850nonRadio}}
\tablehead{
\colhead{ID} &
\colhead{$S_{850}$} & 
\colhead{$S\mathrm{_{radio}}$} & 
\colhead{Hi$z$Idx} &
\colhead{ID} &
\colhead{$S_{850}$} & 
\colhead{$S\mathrm{_{radio}}$} & 
\colhead{Hi$z$Idx} &
\colhead{ID} &
\colhead{$S_{850}$} & 
\colhead{$S\mathrm{_{radio}}$} & 
\colhead{Hi$z$Idx} \\
\colhead{} &
\colhead{(mJy)} &
\colhead{($\mu$Jy)} &
\colhead{(850)} &
\colhead{} &
\colhead{(mJy)} &
\colhead{($\mu$Jy)} &
\colhead{(850)} &
\colhead{} &
\colhead{(mJy)} &
\colhead{($\mu$Jy)} &
\colhead{(850)}
}
\startdata
SD850\_COSMOS\_T1\_A02 & $9.0\pm1.3$ & $<11.5$ & $>0.43$  & A12 & $7.6\pm1.8$ & $<11.5$ & $>0.36$  & A31 & $7.2\pm2.0$ & $<11.5$ & $>0.34$ \\
                   A03 & $11.8\pm1.9$ & $<11.5$ & $\mathbf{>0.56^{*}}$ & A13 & $6.7\pm1.6$ & $<11.5$ & $>0.32$  & B03 & $5.5\pm1.1$ & $<11.5$ & $>0.26$ \\
                   A06 & $9.5\pm2.0$ & $<11.5$ & $>0.45$  & A19 & $5.1\pm1.2$ & $<11.5$ & $>0.24$  & B05 & $8.0\pm1.8$ & $<11.5$ & $>0.38$ \\
                   A07 & $9.5\pm2.0$ & $<11.5$ & $>0.45$  & A22 & $5.5\pm1.4$ & $<11.5$ & $>0.26$  & B07 & $8.5\pm1.9$ & $<11.5$ & $>0.40$ \\
                   A08 & $6.2\pm1.5$ & $<11.5$ & $>0.30$  & A26 & $4.0\pm0.9$ & $<11.5$ & $>0.19$  &     &             &         & \\
                   A10 & $7.9\pm1.9$ & $<11.5$ & $>0.38$  & A29 & $6.3\pm1.8$ & $<11.5$ & $>0.30$  &     &             &         & \\
\hline
   SD850\_EGS\_T1\_A09 & $5.7\pm1.4$ & $<130$  & $>0.04$  & A12 & $4.7\pm1.1$ & $<130$  & $>0.04$  &     &             &         & \\
                   A11 & $7.1\pm1.7$ & $<130$  & $>0.05$  & A13 & $5.0\pm1.3$ & $<130$  & $>0.04$  &     &             &         & \\
\hline
SD850\_GOODSN\_T1\_A01 & $6.7\pm1.4$ & $<20$   & $>0.33$  & A02 & $5.9\pm1.3$ & $<20$   & $>0.30$  &     &             &         & \\
\hline
    SD850\_LH\_T1\_A04 & $4.7\pm1.2$ & $<13.5$ & $>0.35$ \\
\hline
   SD850\_UDS\_T1\_A05 & $7.5\pm1.2$ & $<60$   & $>0.13$  & A45 & $4.4\pm1.1$ & $<60$   & $>0.07$  & A79 & $3.8\pm1.0$ & $<60$   & $>0.06$ \\
                   A08 & $6.7\pm1.2$ & $<60$   & $>0.11$  & A47 & $4.4\pm1.1$ & $<60$   & $>0.07$  & A81 & $3.8\pm1.0$ & $<60$   & $>0.06$ \\
                   A10 & $6.2\pm1.2$ & $<60$   & $>0.10$  & A50 & $4.2\pm1.1$ & $<60$   & $>0.07$  & A82 & $3.9\pm1.1$ & $<60$   & $>0.06$ \\
                   A11 & $5.8\pm1.1$ & $<60$   & $>0.10$  & A51 & $4.7\pm1.1$ & $<60$   & $>0.08$  & A84 & $3.4\pm1.0$ & $<60$   & $>0.06$ \\
                   A15 & $6.0\pm1.2$ & $<60$   & $>0.10$  & A54 & $4.4\pm1.1$ & $<60$   & $>0.07$  & A85 & $3.6\pm1.0$ & $<60$   & $>0.06$ \\
                   A16 & $6.7\pm1.4$ & $<60$   & $>0.11$  & A55 & $4.2\pm1.1$ & $<60$   & $>0.07$  & B14 & $5.7\pm1.2$ & $<60$   & $>0.09$ \\
                   A18 & $5.6\pm1.1$ & $<60$   & $>0.09$  & A56 & $4.1\pm1.0$ & $<60$   & $>0.07$  & B18 & $5.4\pm1.1$ & $<60$   & $>0.09$ \\
                   A19 & $5.9\pm1.1$ & $<60$   & $>0.10$  & A59 & $4.4\pm1.1$ & $<60$   & $>0.07$  & B22 & $5.2\pm1.1$ & $<60$   & $>0.09$ \\
                   A23 & $5.6\pm1.2$ & $<60$   & $>0.09$  & A60 & $4.2\pm1.1$ & $<60$   & $>0.07$  & B24 & $4.6\pm1.0$ & $<60$   & $>0.08$ \\
                   A33 & $4.9\pm1.1$ & $<60$   & $>0.08$  & A61 & $4.2\pm1.0$ & $<60$   & $>0.07$  & B27 & $4.5\pm1.1$ & $<60$   & $>0.08$ \\
                   A34 & $5.0\pm1.1$ & $<60$   & $>0.08$  & A68 & $4.0\pm1.1$ & $<60$   & $>0.07$  & B28 & $4.9\pm1.1$ & $<60$   & $>0.08$ \\
                   A37 & $4.5\pm1.1$ & $<60$   & $>0.08$  & A71 & $3.7\pm1.0$ & $<60$   & $>0.06$  & B32 & $4.5\pm1.1$ & $<60$   & $>0.08$ \\
                   A38 & $4.7\pm1.1$ & $<60$   & $>0.08$  & A72 & $3.8\pm1.0$ & $<60$   & $>0.06$  & B34 & $4.3\pm1.0$ & $<60$   & $>0.07$ \\
                   A39 & $4.5\pm1.1$ & $<60$   & $>0.08$  & A73 & $3.9\pm1.0$ & $<60$   & $>0.06$  & B37 & $4.1\pm1.0$ & $<60$   & $>0.07$ \\
                   A42 & $4.5\pm1.1$ & $<60$   & $>0.07$  & A78 & $3.8\pm1.0$ & $<60$   & $>0.06$  & B42 & $4.0\pm1.1$ & $<60$   & $>0.07$ \\
                   A40 & $4.7\pm1.0$ & $<60$   & $>0.08$  & A80 & $4.6\pm1.3$ & $<60$   & $>0.08$  & B43 & $4.2\pm1.0$ & $<60$   & $>0.07$ \\
\enddata
\tablecomments{Tier 1 SPIRE dropouts covered by various radio surveys but are not detected. 
As in Table \ref{tab:SD850CatSample}, the leading string in the object ID is omitted
for clarity except when the object is the first entry of a given field.
The upper limits of their radio flux densities are based on the radio catalogs described in
\S \ref{sec:SD850_Radio}. The lower limits of Hi$z$Idx(850) are calculated using these upper
limits. The one that has Hi$z$Idx(850)~$\geq 0.5$ (potentially at $z>6$) is boldfaced and
marked with asterisk (see \S \ref{sec:highz}).
}
\end{deluxetable*}

   In summary, the radio data cover most of these fields. About 23\% of the
Tier 1 SPIRE dropouts are detected, and all of them are in the star formation
dominated regime. There are still $\sim$77\% of the Tier 1 SPIRE dropouts not
detected in radio, partly due to the fact that the radio data are still not
deep enough. However, there is also a possibility that some of them are at
high redshifts, which we will discuss further in \S \ref{sec:highz}.

\section{Discussion}

   In the above sections, we present our large samples of 500~$\mu$m risers and
SPIRE dropouts. Admittedly, these samples suffer from the same drawback of the
parent catalogs that we have used (the HerMES DR4 catalogs and the S2CLS
catalog), namely, the difficulty in obtaining high-precision photometry due to
source blending in single-dish data, which is a long-standing problem. We
refer the reader to the relevant documentations cited above for the details
such as source incompleteness, spurious detections, photometric errors, etc. 
Moreover, there could be an additional source of incompleteness introduced in
our visual inspection step, which cannot be assessed through simulations
because we do not have the means to reproduce the parent catalogs. Therefore,
any applications of our samples for statistical purposes should be done with 
caution. With these caveats in mind, here we discuss some limited statistical
properties of our samples, their prospect of being at high $z$, and their 
implications for the dust-embedded star formation in the early universe,
all to the extent that we believe is still reasonable.

\subsection{Surface density of 500~$\mu$m risers \label{sec:500R_density}}

   At the first glance, it seems that the surface density of 500~$\mu$m risers
varies greatly among these 11 fields (see Table 1): the average is 
6.45~deg$^{-2}$ with the dispersion of 2.54~deg$^{-2}$, and the highest value
is a factor of 4.4 of the lowest. We show that, however, this is
largely due to the different sensitivity levels (i.e., survey limits) of the
fields. 

\begin{figure*}[t]
\plotone{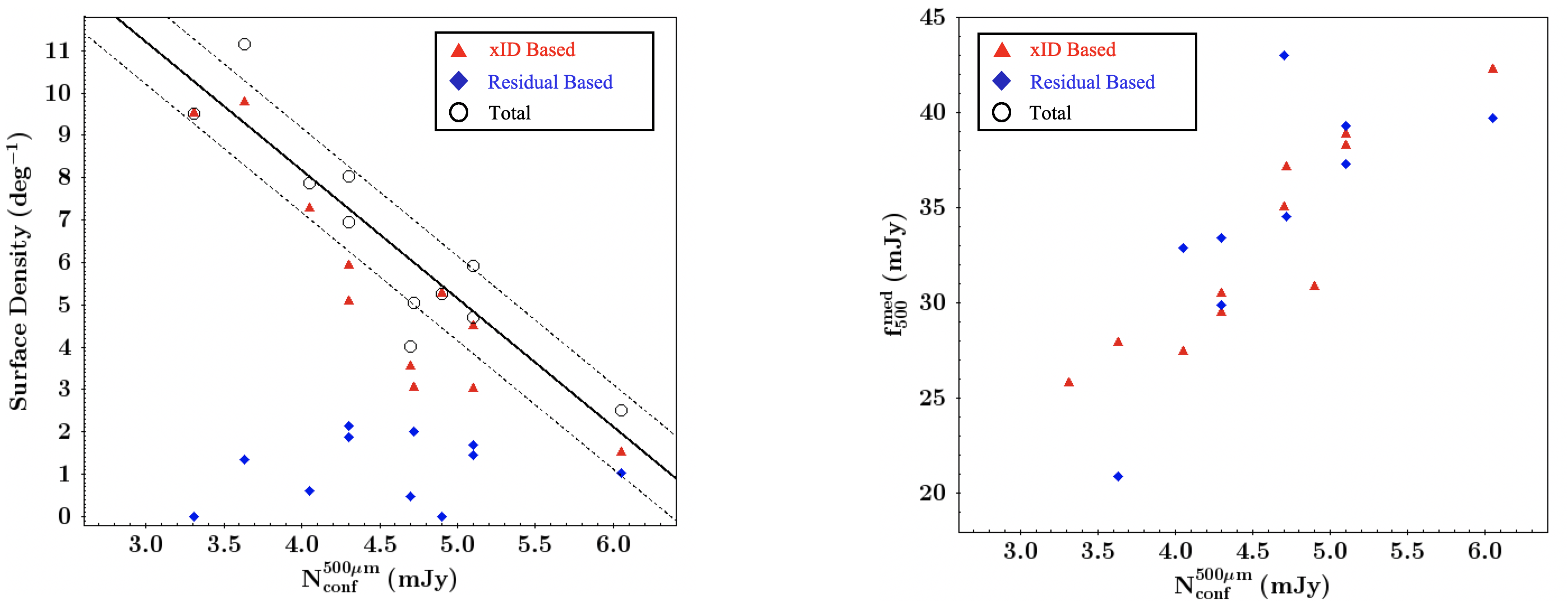}
\caption{(left) Surface densities of the 500~$\mu$m risers in all fields
against the confusion noises in 500~$\mu$m, $\mathrm{N_{conf}^{500\mu m}}$,
which are used as the proxies to the survey limits. The filled red triangles
and the blue diamonds are for the objects selected based on the HerMES DR4 xID 
catalogs and the residual maps, respectively, and the open black circles
indicate the sum of the two.
(right) Median 500~$\mu$m flux densities of the 500~$\mu$m risers against
against $\mathrm{N_{conf}^{500\mu m}}$ in all fields. Legends are the same as
in the left panel. For clarity, the two fields that do not have residual-based
objects are not shown.
}
\label{fig:500umRiserDensityMedFlux}
\end{figure*}

   The survey limits of the HerMES DR4 data are not directly given in the
existing HerMES documentations. For our purposes, we adopt the confusion
noise (see \S 3 and Table \ref{tab:500Rsummary}) as the proxy to the 
sensitivity, because it is the dominant noise term in all of the HerMES data
studied here. The left panel of Figure \ref{fig:500umRiserDensityMedFlux} shows
the surface densities of the 500~$\mu$m risers versus the confusion noises in 
500~$\mu$m ($\mathrm{N_{conf}^{500\mu m}}$). The open circles
indicate the total surface densities in the field, while the red triangles and
the blue diamonds are for the densities calculated using the xID-based and
the residual-based objects, respectively. Clearly, the xID-based surface
density increases as $\mathrm{N_{conf}^{500\mu m}}$ decreases, i.e., as the
sensitivity improves
\footnote{As the SPIRE observations were simultaneous in all three bands, we
see the same trends when using the confusion noise in 250~$\mu$m as the
variable. The 350~$\mu$m total errors as reported in the DR4 xID catalogs are
incorrect in ELAIS-N2, which result in the incorrect calculation of
$\mathrm{N_{conf}^{350\mu m}}$ in the field. After discarding this field, we
also obtain similar trends using $\mathrm{N_{conf}^{350\mu m}}$ as the 
variable. The erroneous $\mathrm{N_{conf}^{350\mu m}}$
do not affect our target selection, however.}.
This indicates that there are more 500~$\mu$m risers at the fainter levels. 
The total surface density shows a similar behavior, because the total sample
is dominated by the xID-based objects. 

   The behavior of the surface density of the residual-based objects, however,
is almost flat. To further investigate this seemingly counterintuitive 
problem, we check the median $f_{500}$ values ($f_{500}^{med}$) of the 
500~$\mu$m risers in all fields. These are shown in the right panel of 
Figure \ref{fig:500umRiserDensityMedFlux} against 
$\mathrm{N_{conf}^{500\mu m}}$ using the same legends. The xID-based objects
show the trend of decreasing $f_{500}^{med}$ with decreasing 
$\mathrm{N_{conf}^{500\mu m}}$, which is consistent with the expectation that
when we reach fainter limits, we find fainter 500~$\mu$m risers. The
residual-based objects also follow the same trend in general, which is to say
that this type of object is also fainter when we reach fainter survey limits.
This suggests that the flat behavior of the blue diamonds in the left panel is
simply due to the canceling of two opposite effects: as the sensitivity
improves, some 500~$\mu$m risers that would not be detected in 250~$\mu$m at a
brighter level are now detected in 250~$\mu$m (they would move from the
``residual-based'' sample to the ``xID-based'' sample); in the meantime, roughly
the same amount of 500~$\mu$m risers not detectable in 500~$\mu$m previously at
a brighter level now becomes detected in 500~$\mu$m (roughly the same number of
objects replenishes the ``residual-based'' sample).

   To better describe the behavior of the open circle in the left panel of
Figure \ref{fig:500umRiserDensityMedFlux}, we perform a linear fit and obtain
$\Sigma = -3.03\times \mathrm{N_{conf}^{500\mu m}} + 20.3$, with the
dispersion of 1.0. 

\subsection{Comparison to the literature}

    Here we compare to the surface densities of 500~$\mu$m risers as inferred
from the work of other teams as mentioned in \S 1. 

    The sample of \citet[][]{Dowell2014} using the early HerMES data consists
of 39 objects over 21~deg$^2$, with $f_{500}^{med}=31.9$~mJy. Therefore, their
overall surface density is 1.9~deg$^{-2}$. Their objects have been
selected using $f_{500}\geq f_{350}\geq f_{250}$ (as measured on their
smoothed maps), which are similar to (but not exactly the same as) our
criteria. They also impose an additional requirement of $f_{500}\geq 30$~mJy.
It is difficult to compare to the S/N level of their objects, because they
require S/N $>4$ on the difference maps. Ignoring this complication, we select
a subsample from our objects, requiring $f_{500}\geq 30$~mJy. This results in
396 objects and implies the surface density of 3.7~deg$^{-2}$, which is 
$2\times$ higher than that of \citet[][]{Dowell2014}.

    The sample of \citet[][]{Asboth2016} in the 274~deg$^2$ HeLMS field 
consists of 477 objects, implying the surface density of 1.7~deg$^{-2}$.
Their objects are selected using the same color criteria as ours but are all
significantly brighter, with $f_{500}>52$~mJy. Our sample only has 39 objects
that have $f_{500}>52$~mJy, and the corresponding surface density would only
be 0.37~deg$^{-2}$, which is a factor of 4.2 lower.

    The sample of 7971 objects of \citet[][]{Ivison2016} are selected using 
$f_{500}/f_{350}\geq 0.85$ and $f_{500}/f_{250}\geq 1.5$, which are less
stringent than ours. As their full sample are not publicly available,
it is difficult to make a direct comparison.  Taken
at the face value, their 7971 objects over 660~deg$^2$ implies a surface
density of 12.1~deg$^{-2}$. They also state that at least $26\pm 5$\% of their
objects are reliable, which then implies the minimum of 3.1~deg$^{-2}$. For
comparison, the overall surface density of our sample is 5.9~deg$^{-2}$.

    The sample of \citet[][]{Donevski2018} consists of 133 objects over
55~deg$^2$, implying 2.4~deg$^{-2}$. They use the same color criteria as ours,
but impose the additional requirements of $f_{500}>30$~mJy (corresponding to
S/N $>4$ on their 500~$\mu$m maps) and $f_{250}>13.2$~mJy. We select a 
subsample from our objects that meet the same flux density cutoffs and obtain
381 objects. This implies 3.6~deg$^{-2}$, which is significantly larger than
theirs.

\subsection{Surface density of SPIRE Dropouts \label{sec:SD850_density}}

   The SPIRE dropouts among the S2CLS fields also seem to vary greatly in 
surface density (see Table \ref{tab:SD850summary}). However, this again is 
largely due to a systematic effect: the selection of our SPIRE dropouts depends
on the sensitivity of the S2CLS 850~$\mu$m data, which varies considerably
among the fields.

   To demonstrate this point, let us take the Tier 1 samples in the UDS and the
COSMOS fields for examples. These two fields contribute 60.0\% (57 objects) and
27.4\% (26 objects) of the total, respectively, and are the only two that have
sufficient statistics for our purpose. While the former covers a smaller area
than the latter one (0.96 versus 1.34~deg$^2$), it has detected more 850~$\mu$m
objects than the latter (1085 vs. 719 objects) because of its better sensitivity
(0.9 versus 1.6 mJy~beam$^{-1}$). This is also reflected in the SPIRE dropouts
that we selected: most of these objects in the UDS field have $S_{850}<5$~mJy
(40 out of 57), while only 3 (out of the 26) in the COSMOS field are in the same
regime.

   According to \citet[][see their Fig. 8]{Geach2017}, the UDS field is 
$\sim 95$\% complete at $S_{850}=5$~mJy, and so the incompleteness
correction is negligible above this threshold. The surface density of its
SPIRE dropouts at $S_{850}\geq 5$~mJy is therefore 17.7~deg$^{-2}$. The COSMOS
field, on the other hand, is only $\sim 53$\% complete at this level, and we
apply the incompleteness corrections to its SPIRE dropout number counts at
$S_{850}\geq 5$~mJy in the successive bins of 0.5~mJy width using Fig. 8 of 
\citet[][]{Geach2017}. The corrected, cumulative number count at
$S_{850}\geq 5$~mJy is 28 (as oppose to the original 23 objects), and the
corresponding surface density is 20.9~deg$^{-2}$. The difference between the
two fields is well within the Poissonian noise. Taking the average of the two,
we obtain the SPIRE dropout surface density of $19.3\pm 1.6$~deg$^{-2}$ at
$S_{850}\geq 5$~mJy.

    We note that the above value is only applicable to degree-size fields.
As the spatial distributions of SPIRE dropouts show a significant clustering
feature (see Figure \ref{fig:SD850Pos}), the surface density can vary 
significantly in small fields. In the EGS (0.32~deg$^2$) and the LH
(0.28~deg$^2$) fields, we have three (out of six total) and two (out of three
total) SPIRE dropouts at $S_{850}\geq 5$~mJy, respectively. The corresponding
surface densities would be 9.4 and 7.1~deg$^{-2}$, respectively, which are both
much lower than the value derived above. On the contrary, in the GOODS-N field
that covers only 0.07~deg$^2$, all the three SPIRE dropouts are at
$S_{850}\geq 5$~mJy. This would imply the surface density of 42.9~deg$^{-2}$,
which is $2.2\times$ higher than the average.

\subsection{Prospects of 500~$\mu$m risers and SPIRE Dropouts Being at High z
\label{sec:highz}}

   While the initial motivation of our work is to select high-$z$ ULIRGs in
the EoR (see \S 1), it is extremely difficult to further purify our 
500~$\mu$m riser and SPIRE dropout samples to achieve this goal. Without
knowing the exact location of the dust emission peak, the 500~$\mu$m riser
method is only expected to select objects at $z\gtrsim 4$ in general but not to
create a sample falling within a well-defined redshift range. It is also 
expected that the selection could be severely contaminated by objects at
$z\approx 3$ due to the degeneracy between dust temperature and redshift. 
For example, \citet[][]{Asboth2016} carried out 3 mm spectroscopy for two
of their sources and find that one is at $z=5.162$ and the other is at
$z=3.798$. \citet[][]{Ivison2016} derived photometric redshifts for
their 500~$\mu$m risers based on the FIR/sub-mm SEDs, and concluded that the 
median redshift is at 3.66 and that only a third of their 500~$\mu$m risers
lie at $z>4$. This is partly confirmed by \citet[][]{Fudamoto2017}, who have
done 3 mm spectroscopy on a small subsample of 17 very red objects in
\citet[][]{Ivison2016}. Among the seven objects that have secure redshifts, one
is at $z=3.8847$ and five are at $4<z<5$. Encouragingly, there is one object at
$z=6.027$ \citep[see also][]{Zavala2018}. However, this object is not the 
reddest in their sample, which demonstrates the large uncertainty in 
estimating redshifts for 500~$\mu$m risers. This is also true for 850/870~$\mu$m
risers. For example, \citet[][]{Oteo2018} reported a 870~$\mu$m riser; however,
it is only at $z=4.002$. The situation for the SPIRE dropouts is even more
complicated because we have even less information about their SEDs. 
Nevertheless, a small fraction of our objects do have further constraints from
the radio data, which we discuss in detail below.

\subsubsection{Constraints from Radio Data}

   For the objects that fall within the coverage of radio observations, it could
be possible to further constrain their redshifts by incorporating the radio
data.  A number of photometric redshift derivation schemes have been proposed
based on the FIR-radio relation \citep[see e.g.,][]{CarilliYun1999, Barger2000}.
However, such methods all suffer from a couple of major caveats. First, one
would need to assume a fixed power-law index for the radio emission, which could
have a wide range of uncertainty over $\alpha \sim -0.3$ to $-0.8$. Second,
one would also need to assume a fixed dust temperature, which could have a wide
spread of $\sim 20$--60~K. While the choices of these parameters could be
fine-tuned to work reasonably well within the typical redshift range of 
SMGs at $z\approx 2$--4, it is unclear how they would perform beyond this range.
For example, the famous HDF 850.1, the brightest 850~$\mu$m source in the
Hubble Deep Field with $S_{850}=7.0\pm0.4$~mJy \citep[][]{Hughes1998}, has
$S_{1.4}=16.73\pm 4.25$~$\mu$Jy \citep[][]{Cowie2009}. Using the formalism of
\citet[][]{CarilliYun1999}, one would obtain $z_{ph}=4.6$. If using that of
\citet[][]{Barger2000}, one would obtain $z_{ph}=3.8$. Both are significantly
different from the spectroscopic redshift of $z=5.183$ \citep[][]{Walter2012}.
Another example is HFLS3 at $z=6.34$ of \citet[][]{Riechers2013}, which
has $S_{880}=33.0\pm 2.4$~mJy (the SMA measurement at 880~$\mu$m) and 
$S_{1.4}=59\pm 11$~$\mu$Jy. Ignoring the small difference between $S_{880}$ and
$S_{850}$, one would then obtain $z_{ph}=5.0$ and 4.1 based on
\citet[][]{CarilliYun1999} and \citet[][]{Barger2000}, respectively, both of
which are also significantly different from the true value.

  For this reason, we do not attempt in this work to directly derive $z_{ph}$
using the radio data. Instead, we only select the objects that are radio weak
as the most promising candidates at $z>6$, among which some could be in the
EoR. 

  The problem is how to define ``radio weak'' in this context. Lacking anything
else being a better choice, we use the aforementioned HDF 850.1 and HFLS3 
as the references. Recall that HDF 850.1, or our \texttt{SD850\_HDFN\_T1\_A02}, 
is a SPIRE dropout, while FLS 3, or our \texttt{500R\_FLS\_T1\_x44}, is a
500~$\mu$m riser (but not an 850~$\mu$m riser). For the sake of the argument, we
introduce a ``high-z index,'' or ``Hi$z$Idx'' for short, to quantify an object
being radio weak by using the flux density ratio between the FIR/submm and 
radio:

\begin{equation}\label{eq:highzindex}
    {\mathrm{Hi}}z{\mathrm{Idx}} =
    \begin{cases}
       f_{500}\times 10^{-3}/S_{1.4}\, & (500~\mu m\ {\mathrm{risers}}) \\
       S_{850}\times 10^{-3}/S_{1.4}\, & ({\mathrm{SPIRE\ dropouts}})
    \end{cases}
\end{equation}

   The 850~$\mu$m photometry for HDF 850.1 (at $z=5.18$) is somewhat different 
among the data published by different groups
\citep[][]{Walter2012}\citep[see also][]{Cowie2017}.
Here we adopt $S_{850}=5.9\pm1.3$~mJy from the S2CLS catalog and obtain
Hi$z$Idx(850)~$=0.35$. For FLS 3 (at $z=6.43$), we obtain Hi$z$Idx(500)$=0.75$
and Hi$z$Idx(850)~$=0.56$, respectively.

   Using the above numbers as a guide, we adopt Hi$z$Idx(500)~$\geq 0.7$ to 
further select 500~$\mu$m risers at $z>6$. To select objects at $z>6$ from 
SPIRE dropouts, we adopt Hi$z$Idx(850)~$\geq 0.5$. We note that these ad hoc
thresholds are so chosen only for the sake of the argument here and are the
best that one could adopt at this stage. Other better choices will have to wait
until spectroscopic verifications in the future.

\subsubsection{High-z candidates in the 500~$\mu$m Riser Sample}

  For the 500~$\mu$m risers that have radio counterparts (as described in 
Section 5.4), we calculate their Hi$z$Idx(500).
For those that are within the radio coverage but are not detected, we calculate
the lower limits of Hi$z$Idx(500) using the upper limits of the radio
flux densities appropriate in the fields. For the radio observations that are
not made in 1.4~GHz, we convert the result to $S_{1.4}$ by assuming
$S(\nu)\propto \nu^{-0.8}$. These results are listed in
Table \ref{tab:500RtoDeepRadio} and \ref{tab:500RnonRadio}.

  In total, 19 of our 500~$\mu$m risers have Hi$z$Idx(500)~$\geq 0.7$,
distributed over five fields. The COSMOS field is the only one that offers
sufficient statistics over a large area to allow a meaningful estimate of the
surface density for such radio-constrained high-$z$ candidates. Recall that it
has 17 500~$\mu$m risers within the 3~GHz coverage, 10 of which have 3~GHz
counterparts. Four of these 10 have Hi$z$Idx(500)~$\geq 0.7$. The seven that
are not detected at 5~$\sigma$ all have Hi$z$Idx(500)~$\geq0.7$. Therefore,
there are 11 candidates (or 64.7\% out of the total of 17) at $z>6$ over
2~deg$^2$ satisfying our criterion, which correspond to a density of 
5.5~deg$^{-2}$.

\subsubsection{High-z candidates in the SPIRE Dropout Sample}

   Similarly, we calculate Hi$z$Idx(850) or the lower limits for the 
SPIRE dropouts and list the results in the catalog. As it turns out, none of
the objects that have identified radio counterparts satisfy our criterion for
$z>6$. However, we can still obtain useful information from those that are not
detected in the existing radio data.

   The most stringent limits are again from the COSMOS field over 1.34~deg$^2$,
which has 16 SPIRE dropouts that do not have counterparts in the 3~GHz catalog.
Adopting the 5~$\sigma$ limit of 11.5~$\mu$Jy and assuming
$S(\nu)\propto \nu^{-0.8}$ as before, we find that one object, 
\texttt{SD850\_COSMOS\_T1\_A03} with $S_{850}=11.8\pm1.9$~mJy,
has Hi$z$Idx(850)~$>0.56$ and thus satisfies our criterion for $z>6$. Therefore,
we can derive the lower limit of $z>6$ objects among the SPIRE dropouts to be
0.8~deg$^{-2}$. The upper limit can be obtained by
assuming that all these 16 objects are not detected even at higher
sensitivity levels. Indeed, if they all remain radio undetected at the
$\sim 2.5$~$\sigma$ level, they would all have Hi$z$Idx(850)~$>0.5$. Out of
these 16 objects, 15 have $S_{850}\geq 5.0$~mJy, and therefore we derive
the upper limit of the surface density as 11.2~deg$^{-2}$ to 
$S_{850}=5.0$~mJy. If we further consider the incompleteness correction for
the COSMOS field in 850~$\mu$m (see in \S 7.3), the upper limit is
13.6~deg$^{-2}$.

   Following the same methodology, we check the upper limits inferred from
the other fields, bearing in mind that they have highly nonuniform
sensitivities in their radio data. On the order of the field size, the UDS
(0.96~deg$^2$), the EGS (only about half, or $\sim 0.16$~deg$^2$ covered by the
radio data), the LH (0.28~deg$^2$), and the GOODS (0.07~deg$^2$) fields have
13, 2, 1, and 1 radio-undetected objects at $S_{850}\geq 5$~mJy, respectively.
The corresponding upper limits for $z>6$ objects are therefore 13.5, 12.5, 3.6,
and 14.3~deg$^{-2}$ to $S_{850}=5.0$~mJy.

   We note in passing that the SPIRE dropout of \citet[][]{Greenslade2019}, 
their ``NGP6\_D1,'' does not meet our Hi$z$Idx criterion for $z>6$. This 
source has $S_{850}=12.3\pm2.5$~mJy. Their VLA 6~GHz 
measurement gives $S_{6.0}=16.9\pm4$~$\mu$Jy, which corresponds to
$S_{1.4}=54.1$~$\mu$Jy if assuming $S(\nu)\propto \nu^{-0.8}$. This gives 
Hi$z$Idx(850)=0.23. To make Hi$z$Idx(850) $>0.7$, it would have to have a
very flat SED slope of $\alpha\leq 0.03$, which does not seem likely.

\subsubsection{Global star formation rate density at $z>6$ hidden by dust}

   Using the results in \S 7.4.2 and 7.4.3, here we attempt to estimate the
IR-based (i.e., dust-embedded) global star formation rate density (GSFRD; 
$\dot{\rho_{\ast}}$(IR)) at $z>6$. 
We only use the values based on the COSMOS field. For the surface density of
$z>6$ SPIRE dropouts, we adopt the average of the lower and the upper limits in
Section 6.4.3 as the representative value, i.e., 7.2~deg$^{-2}$. The combined
surface density of 500~$\mu$m risers and SPIRE dropouts at $z>6$ is therefore
12.7~deg$^{-2}$. We assume that this is applicable within $6<z<7$, which means
that the volume density is $1.4\times 10^{-6}$~Mpc$^{-3}$.

   For practical purposes, we further assume that these objects all have
$L_{IR}=10^{13}L_\odot$, 
which is likely to the lower end of the true
distribution. For example, HFLS3, which is our \texttt{500R\_FLS\_T1\_x44},
has $L_{IR}=1.55\times10^{13}L_\odot$ even after the correction for the 
gravitational lensing \citep[][]{Cooray2014}.
We apply the conversion given by 
\citet[][]{Kennicutt1998}, which is $\mathrm{SFR}=1.0\times10^{-10}L_{IR}$
after adjusting for a Chabrier initial mass function \citep[][]{Chabrier2003}.
This translates to $\mathrm{SFR}=10^{3}M_\odot$~yr$^{-1}$ per object.
Therefore, we obtain 
$\dot{\rho_{\ast}}\mathrm{(IR)}=1.4\times 10^{-3}M_\odot$~yr$^{-1}$~Mpc$^{-3}$.
This is comparable to the UV-based GSFRD derived using Lyman-break galaxies
(LBGs). For example, \citet[][]{Yan2010} have obtained 
$\dot{\rho_{\ast}}\mathrm{(UV)}=(12.33, 5.50)\times 10^{-3}$~$M_\odot$~yr$^{-1}$~Mpc$^{-3}$ at $z=(6.0, 7.0)$, respectively. 
Note that these 500~$\mu$m risers and SPIRE dropouts are not likely visible in
the rest-frame UV and hence are not likely to manifest themselves as LBGs
(HFLS3 is an example),
which means that $\dot{\rho_{\ast}}\mathrm{(UV)}$ after dust-reddening
correction cannot account for $\dot{\rho_{\ast}}\mathrm{(IR)}$. In other words,
the GSFRD value at high redshifts is likely significantly higher than that
derived based on LBGs alone. This point has been raised before
\citep[see, e.g.][]{Cowie2009}, and recently was further reinforced by
\citet[][]{Wang2019} in their 870~$\mu$m detections of H dropouts (3.6~$\mu$m 
sources that are invisible from UV to $H$ band), the bulk of which are likely
at $z>3$ and some of which could be at $z>6$. \citet[][]{Wang2019} estimated
that the volume density of their objects is $2\times 10^{-5}$~Mpc$^{-3}$,
which is about an order of magnitude higher than that of our objects. However,
as their $L_{IR}$ are at least an order of magnitude lower, the total
contribution to the GSFRD from these less luminous objects 
($0.9\times 10^{-3}M_\odot$~yr$^{-1}$~Mpc$^{-3}$ at $z=6$) is comparable to
that from our objects.

   We shall caution that our conclusions above hinge upon a number of
assumptions, most importantly the validity of the $z>6$ candidates derived
based on Hi$z$Idx. Obviously, spectroscopic confirmation of a significant
sample like what presented here is the only path to solving the problem.

\section{Summary}

   We have carried out a comprehensive search for 500~$\mu$m risers using
the HerMES data in 106.5~deg$^2$ and for SPIRE dropouts using the S2CLS data
in the 2.98~deg$^2$ that overlap the HerMES coverage. Both types of objects are 
candidate ULIRGs at high redshifts, some of which could be at $z>6$ and in 
the EoR.  The main objective of this paper is to present the selections and the
catalogs of these rare sources to facilitate follow-up observations.
We focus on the ``Tier 1'' objects, which we believe are the least affected by 
contaminations and thus are valuable targets for future studies. For 
completeness, the catalogs of the less promising, ``Tier 2'' objects are also
included in this paper. 

   In total, we present 629 Tier 1 500~$\mu$m risers in 11 HerMES fields, 
all detected at $\mathrm{S/N\geq 5}$ (confusion noise included) in 500~$\mu$m.
These objects satisfy the simple criterion of $f_{500}>f_{350}>f_{250}$, which
is similar to (but not exactly the same as) most of the existing selections of
500~$\mu$m risers in the literature. Compared to the existing ones, our sample
is unique in its flux limit and spatial coverage. About 77.4\% of this sample
(487 objects) are selected from the HerMES DR4 xID catalogs (``xID based''; the
objects have reported 250~$\mu$m detections), and the other 22.6\% (142 objects)
are selected through detecting objects in the 500~$\mu$m residual maps generated
by subtracting off the sources in the xID catalogs (``residual based''; the 
objects do not have reported 250~$\mu$m detections). Their spatial 
distributions show obvious patterns of clustering. The inferred surface density
depends on the survey limit, which ranges from 
$\sim 2.1$~\si{\per\deg\squared} at the confusion noise level of 
$\mathrm{N_{conf}^{500\mu m}}\approx 6$~mJy (or equivalently, 5~$\sigma$
flux density limit of $f_{500}\gtrsim 30$~mJy) to 
$\sim 8.2$~\si{\per\deg\squared} at
$\mathrm{N_{conf}^{500\mu m}}\approx 4$~mJy (or equivalently, 5~$\sigma$
limit of $f_{500}\gtrsim 20$~mJy).

   The search for SPIRE dropouts is relatively new, and in fact our sample is
the only significant sample of its kind to date. Our Tier 1 catalog includes 95 
SPIRE dropouts in five S2CLS fields that also have HerMES SPIRE data. While 
the limit of the SPIRE data cannot guarantee that these SPIRE dropouts are 
850~$\mu$m risers, any 850~$\mu$m risers in these fields must be among these 
SPIRE dropouts. They all have ``detection\_SNR''~$\geq 5$ in the S2CLS catalog,
and hence are highly reliable detections in 850~$\mu$m. About 76.8\% of them 
(73 objects) do not have an entry in the HerMES xID catalog (selected by
``Method A''). The other 23.2\% (22 objects) have $\mathrm{S/N<3}$ in the xID
catalog (selected by ``Method B'') and thus could actually be nondetections in
SPIRE. The ``B'' objects also satisfy
$(S_{850}+0.5\times \mathrm{ErrSr}_{850})>(f_{250}-2\times et_{250})$.
The inferred surface density depends on the survey limits of both S2CLS and
HerMES, and the former dominates. From the results in the COSMOS and the
UDS fields (the only two degree-size fields), we derive their surface
density of $19.3\pm 1.6$~deg$^{-2}$ at $S_{850}\geq 5$~mJy.

   The SEDs alone cannot be used to constrain the redshifts of these objects.
As FIR/sub-mm galaxies at high redshifts should generally be radio weak,
we examine the radio properties of our 500~$\mu$m risers and SPIRE dropouts
using public radio data. Cross-matching with the FIRST data in the northern and
the equatorial fields has revealed seven very strong AGNs among the
500~$\mu$m risers (out of 381 total in these fields), four of which are
(likely) blazars because the rising trend of their SEDs extends into the radio
regime. Deeper radio data are available and cover 42.4\% of the 500~$\mu$m 
risers (267 out of 629 objects) and almost all of the SPIRE dropouts (93 out of
95 objects). The sensitivity limits of these radio data vary greatly across the
fields, resulting in highly different radio detection rates. Nevertheless, the
vast majority of our objects are at the sub-millijansky level and weaker, 
supporting the fact that they are mostly driven by star formation. For ease of
use, the relevant radio properties of these objects are included in the
catalogs. Furthermore, we introduce the Hi$z$Idx parameter, which is the flux
density ratio between FIR/sub-mm and radio. Using the well-studied HDF 850.1
($z=5.18$) and HFLS3 ($z=6.43$) as guides, we propose Hi$z$Idx(500)~$\geq 0.7$
and Hi$z$Idx(850)~$\geq 0.5$ to select $z>6$ objects from the 500~$\mu$m risers
and the SPIRE dropouts, respectively. The COSMOS field is the only area where
the radio data are deep and wide enough to provide stringent constrains on the 
surface density of potential $z>6$ far-IR/sub-mm sources. Based on the 
aforementioned Hi$z$Idx criteria, we find that 19 500~$\mu$m risers and one
SPIRE dropout could be at $z>6$, and we derive the surface density of $z>6$
objects as 5.5~deg$^{-2}$ among the 500~$\mu$m risers and 0.8--13.6~deg$^{-2}$
among the SPIRE dropouts, respectively. Finally, we point out that
dust-embedded GSFRD at $z>6$ could be comparable to that derived based on
LBGs alone. To really understand the role of dust-embedded star formation in
the early universe, however, spectroscopic identification of a significant
$z>6$ sample such as presented here will be necessary.

\acknowledgments{
We would like to thank the anonymous referee for the helpful comments.
This research is supported in part by the National Natural Science Foundation
of China (NSFC, grant No. 11728306), and is also sponsored in part by the 
Chinese Academy of Sciences (CAS), through a grant to the CAS South America
Center for Astronomy (CASSACA) in Santiago, Chile. L.F. acknowledges the support
from the NSFC (grant Nos. 11822303 and 11773020) and Shandong Provincial
Natural Science Foundation, China (ZR2017QA001, JQ201801).
This research has made use of the data from the HerMES project 
(\url{http://hermes.sussex.ac.uk}). HerMES is a Herschel Key Programme 
utilizing Guaranteed Time from the SPIRE instrument team, ESAC scientists, and a
mission scientist.  The HerMES data were accessed through the Herschel Database 
in Marseille (HeDaM; \url{http://hedam.lam.fr}) operated by CeSAM and hosted by
the Laboratoire d'Astrophysique de Marseille. This research has also made use
of the data from the S2CLS program, which was carried out at the James Clerk
Maxwell Telescope (JCMT). The JCMT is operated by the East Asian Observatory on
behalf of The National Astronomical Observatory of Japan; Academia Sinica 
Institute of Astronomy and Astrophysics; the Korea Astronomy and Space Science
Institute; and Center for Astronomical Mega-Science (as well as the National Key
R\&D Program of China with No. 2017YFA0402700). Additional funding support is
provided by the Science and Technology Facilities Council of the United Kingdom
and participating universities in the United Kingdom and Canada.
}

\bibliographystyle{apj.bst}

\appendix

\section{Properties of 500~$\mu$m risers at Submillimeter/Millimeter \label{sec:500R_mmsubmm}}

   Only a small number of our 500~$\mu$m risers fall within the limited S2CLS
coverage. In the S2CLS GOODS-N and LH fields, there are no 500~$\mu$m risers. 
In the COSMOS field, five of our 500~$\mu$m risers are covered, namely,
{\tt 500R\_COSMOS\_T1\_x01}, {\tt x13}, {\tt x21}, {\tt x25}, and {\tt x34}.
Only {\tt x25} is clearly not detected, which we assign the 2~$\sigma$ upper 
limit of $S_{850}<3.2$~mJy (visual examination of the 3~GHz map also shows no
hints of any detection). Among those that are detected, {\tt x21} is matched to
COS.0345; however, its deboosted $S_{850}$ value is reported in the S2CLS 
catalog as $2.80\pm 2.45$~mJy, which is very different from the ``observed''
$S_{850}=7.93\pm 1.90$~mJy and hence is likely erroneous. Here we adopt the 
mean difference between the observed and the deboosted values in this field as
1.54~mJy, and get $S_{850}=6.39$~mJy. The case for {\tt x34} is complicated.
Its centroid is in between two S2CLS objects, COS.0170 and 0191, and is
separated from them by 8.\arcsec2 and 10.\arcsec9, respectively. These two S2CLS
objects have deboosted $S_{850}=5.4\pm1.2$ and $5.0\pm1.3$~mJy, respectively,
and are separated by 17.\arcsec1. It is unclear if {\tt x34} is related with
either, or if it is the blended result of the two. For our practical purposes
here, we assume the latter and further assume that they are at the same 
redshift. By combining these two objects, we get the deboosted 
$S_{850}=10.35\pm 1.80$~mJy.

   In the XMM-LSS field, one of our sources, {\tt 500R\_XMMLSS\_T1\_x70}, is
covered. It actually corresponds to two objects in the S2LS catalog, 
S2CLS UDS.0074 and 0750, which are 12.\arcsec 7 apart and
have deboosted $S_{850}=6.44\pm 1.26$ and $2.61\pm 0.89$ mJy, respectively.
As in the case for {\tt 500R\_COSMOS\_T1\_x34} above, we combine these two
entries and obtain its deboosted $S_{850}=9.05\pm 1.54$~mJy.

  The S2CLS EGS field covers one of our 500~$\mu$m risers, 
{\tt 500R\_EGS\_T1\_x42}, which is matched with EGS.0007 and has 
$S_{850}=7.28\pm 0.91$~mJy in the S2CLS catalog.

   Three of our objects in the ADFS field, {\tt 500R\_ADFS\_T1\_x14}, {\tt x20}
and {\tt x28}, are covered by the AzTEC 1.1~mm observation of
\citet[][]{Hatsukade2011} and are detected. These sources are matched with 
their cataloged objects AzTEC 27, 4, and 170, respectively.

   The above results are summarized in Table \ref{tab:500R_submm_mm}.

\begin{table*}
\caption{Properties of 500~$\mu$m risers in sub-mm/mm
   \label{tab:500R_submm_mm}
}
\begin{ctabular}{lccccc}
\toprule
   Object Name            & $f_{250}$  & $f_{350}$  &  $f_{500}$  & $S_{850}$  &  Hi$z$Idx(500) \\
\midrule
   500R\_COSMOS\_T1\_x01  & $18.1\pm3.4$ & $24.9\pm6.6$ & $26.2\pm5.1$ & $ 7.1\pm1.9$ &  0.56    \\
   500R\_COSMOS\_T1\_x13  & $23.9\pm3.4$ & $28.4\pm6.6$ & $28.6\pm5.1$ & $11.5\pm2.0$ &  0.28    \\  
   500R\_COSMOS\_T1\_x21  & $25.6\pm3.4$ & $30.7\pm6.6$ & $30.8\pm5.0$ & $ 6.4\pm1.9$ &  $>1.45$ \\
   500R\_COSMOS\_T1\_x34  & $31.4\pm3.4$ & $35.0\pm6.6$ & $39.8\pm5.1$ & $10.4\pm1.8$ &  $>1.87$ \\
   500R\_COSMOS\_T1\_x25  & $13.8\pm3.4$ & $21.8\pm6.6$ & $31.4\pm5.3$ &   $<3.2$     &  0.37    \\
   500R\_EGS\_T1\_x42     & $16.0\pm3.5$ & $22.8\pm4.1$ & $27.1\pm4.5$ & $ 7.3\pm0.9$ &  $>0.21$ \\
   500R\_XMMLSS\_T1\_x70 & $29.8\pm4.6$ & $35.1\pm7.7$ & $39.4\pm6.5$ & $ 9.1\pm1.5$ &  $>0.39$ \\
                          &              &              &              &              &          \\
   Object Name            & $f_{250}$  & $f_{350}$  &  $f_{500}$  & $S_{1.1}$  &  Hi$z$Idx(500)   \\
\midrule
   500R\_ADFS\_T1\_x14    & $26.3\pm4.8$ & $29.4\pm5.6$ & $29.6\pm5.6$ & $3.4\pm0.4$  & $>0.03$ \\
   500R\_ADFS\_T1\_x20    & $24.9\pm4.8$ & $35.5\pm5.6$ & $37.1\pm5.7$ & $6.2\pm0.5$  & $>0.04$ \\
   500R\_ADFS\_T1\_x28    & $27.3\pm4.8$ & $32.1\pm5.7$ & $34.0\pm5.8$ & $6.1\pm0.8$  & $>0.03$ \\
\bottomrule
\end{ctabular}
\tablecomments{Sub-mm and mm properties of the 500~$\mu$m risers that have 
SCUBA2 850~$\mu$m or AzTEC 1.1 mm observations (see \S \ref{sec:500R_mmsubmm}).
The complications in the cases of {\tt 500R\_COSMOS\_T1\_x21}, {\tt 500R\_COSMOS\_T1\_x34}
and {\tt 500R\_XMMLSS\_T1\_x70} are discussed in the text.
Hi$z$Idx(500) is defined by Equation \ref{eq:highzindex}, and the values are taken from
Table \ref{tab:500RtoDeepRadio} and \ref{tab:500RnonRadio}.
}
\end{table*}

\section{Full catalogs of 500~$\mu$m risers and SPIRE dropouts}

   The Tier 1 catalogs of 500~$\mu$m risers and SPIRE dropouts are given as
tables below. Their Tier 2 catalogs are provided as online data.
The COSMOS and the GOODS-N fields do not have residual-based 500~$\mu$m risers
(the ``R'' objects) in either Tier 1 or Tier 2. The EGS and the ELAISN1 fields
also do not have residual-based 500~$\mu$m risers in Tier 2. The GOODS-N field
does not have any Tier 2 SPIRE dropouts.

   All catalogs can be obtained as machine-readable tables at the ApJS site, 
and are also available from the authors upon request (yanha@missouri.edu).

\startlongtable
\begin{longrotatetable}


\end{document}